\documentclass[a4paper, 11pt]{article}
\usepackage[utf8]{inputenc}
\RequirePackage[OT1]{fontenc}
\usepackage{amssymb,amsmath,amsfonts,subfig}
\usepackage{caption}
\usepackage{epsfig,latexsym,graphicx}
\RequirePackage[numbers,sort,compress]{natbib}
\usepackage{setspace} 
\usepackage{adjustbox}
\usepackage{float}
\usepackage{multirow}
\usepackage{hyperref}
\usepackage{lipsum}
\usepackage{chngcntr}
\usepackage[nottoc]{tocbibind}
\usepackage[toc,page]{appendix}
\usepackage{longtable}
\usepackage{setspace,color}
\newcommand{\cbl}{\color{black}}

\newtheorem{theorem}{Theorem}[section]

\newtheorem{corollary}[theorem]{Corollary}
\newtheorem{remark}{Remark}[section]

\numberwithin{equation}{section}
\providecommand{\keywords}[1]
{
	\small	
	\textbf{\textit{Keywords:~}} #1
}
\usepackage{setspace}
\begin{document}

\title{\bf{Robust Estimation of Average Treatment Effects from Panel Data}}
\author{
	Sayoni Roychowdhury$^\dagger$, Indrila Ganguly$^\ddagger$ and Abhik Ghosh$^{\dagger\ast}$\\
	$^\dagger$ Indian Statistical Institute, Kolkata, India. \\
	$^\ddagger$ North Carolina State University, Raleigh, NC, USA. \\
	$^\ast$Corresponding author: abhik.ghosh@isical.ac.in
	%{Interdisciplinary Statistical Research Unit\\
	%	Indian Statistical Institute, Kolkata\\
	%	203, B. T. Road, Kolkata-700108\\
	%	West Bengal, India}
}%\\ 
\date{}
\maketitle

%\vspace{4cm}
\begin{abstract}
In order to evaluate the impact of a policy intervention on a group of units over time, 
it is important to correctly estimate the average treatment effect (ATE) measure. 
Due to lack of robustness of the existing procedures of estimating ATE from panel data, in this paper, 
we introduce a robust estimator of the ATE and the subsequent inference procedures using the popular approach of minimum density power divergence inference. 
Asymptotic properties of the proposed ATE estimator are derived and used to construct robust test statistics for testing parametric hypotheses related to the ATE.
Besides asymptotic analyses of efficiency and powers, extensive simulation studies are conducted to study the finite-sample performances of 
our proposed  estimation and testing procedures under both pure and contaminated data. 
The robustness of the ATE estimator is further investigated theoretically through the influence functions analyses.
%Subsequently, based on these influence analyses, another slightly modified robust ATE estimator is proposed 
%to successfully handle data contamination in post-treatment periods, along with empirical illustrations of its robustness.   
Finally our proposal is applied to study the long-term economic effects of the 2004 Indian Ocean earthquake and tsunami
on the (per-capita) gross domestic products (GDP) of five mostly affected countries, namely Indonesia, Sri Lanka, Thailand, India and Maldives. 
\end{abstract} 
%\vspace{2cm}


%\newpage
%\tableofcontents

%\newpage
\section{Introduction}
\label{SEC:intro}

The problem of estimating average treatment effects (ATE) from non-randomized studies has become quite intriguing to economists and social scientists in recent times. 
Although the term ``\textit{treatment}" was initially  associated with the fields of medicine and agriculture, with the passage of time, 
it is being used in a more general sense where the effects of any specific experimentation, event or policy intervention are collectively termed as 
``\textit{treatment}" effects. Impact of any such treatment is statistically estimated via the ATE across several domains of applications
including, among others, economics, business and management sciences, social or political sciences, marketing, etc. 
For example,  the ATE has been used to study the impact of sovereignty change and the CEPA implementation between mainland China and Hong Kong 
(starting in the first quarter of 2004) on the economic growth of Hong Kong in \cite{Hsiao/etc:2012}. 
The impact of opening a new showroom of eyewear brand on its sales has been investigated in \cite{Li:2018}.
Further applications of the ATE can be found in, among many others, \cite{Bai/etc:2014,Ouyang/Peng:2015,Du/Zhang:2015}.

As the name suggests, the ATE is intuitively computed by comparing the outcomes of units in the treatment group to those of units in the control group
and is often expressed as the difference in their mean outcomes.  
But, such a simple approach can yield the true treatment effect only if the two groups are perfectly homogeneous in every possible aspect, 
which is often not the case in practice (not even in many planned randomized experiments if the sample sizes are not large enough).
As put forward by \citet{Sakaguchi:2020}, evaluation of policy (treatment) effects from non-randomized studies generally assumes that, 
given observed covariates, treatment assignments are independent of possible outcomes, which is often violated due to unobserved confounding variables. 
The main challenge to get correct treatment effect from such real-life (non-randomized) data, in fact, 
lies in estimating the counterfactual, the hypothetical outcome of the treated unit in the absence of treatment, 
with utmost accuracy via a proper adjustment for the confounding effects.  

To outmanoeuvre the issue of unobserved confounding variables, suitable latent factor models are proposed to estimate the ATE from panel data 
containing multi-period observations. In any policy intervention (or event), which affects individuals or the society over time,
panel data are the most common type available with some observations being before the intervention event and some being  after it. 
So, it is important to develop appropriate methods for estimating ATE from panel data  and, as expected, 
there are already various methods available for this purpose. 
\citet{Abadie:2005} proposed an innovative Difference-in-Differences (DID) approach for estimating the counterfactual with time-varying latent factors,
which leads to  a consistent estimator of the ATE even when the number of time periods is small provided the numbers of control and treatment units are large. 
However, there are some important limitations of this approach as it assumes no sample selection effect
and  that the average outcomes for the treatment and control units follow \textit{parallel paths} over time in the absence of treatment. 
\citet{Li/Bell:2017a} proposed the Augmented DID (ADID) approach for consistent estimation of the ATE 
using a linear regression of treatment units on the control units over the pre-treatment period; 
%. In this method, the treatment unit is regressed on the control units in the pre-treatment period and the regression coefficients are estimated using least squares. Using these estimates, we estimate the outcome for the treatment unit in the post-treatment period. A simple difference between the actual outcome and estimated outcome for the treatment unit averaged over the post-treatment time period gives the ATE.  The Augmeted DID estimator is not only straightforward to implement, but also 
the resulting estimator is robust to the selection method for non-treated units and also performs well when the \textit{parallel path} assumption is violated.  
For the cases exhibiting nonlinear and/or stochastic trends, \citet{Abadie/Gardeazabal:2003} proposed the synthetic control method (SCM) 
which uses a weighted average of control units to approximate the counterfactual outcome of the treated units in the absence of treatment; 
the weights are restricted to be non-negative and to sum to one. 
%The synthetic control method includes the Difference-in-Differences method as a special case when each control unit is assigned a weight of $1/N_{co}$, the number of control units. 
The performance of SCM relies crucially on the assumption that the weighted average of the control units' sample path is 
parallel to the treated unit's sample path in the absence of treatment. 
If this assumption is violated in practice, \citet{Doudchenko/Imbens:2016} suggested a modified SCM (MSCM) by relaxing the restriction that weights sum to one. 
This modification effectively relaxes the original \textit{parallel line}  assumption to a weaker condition that 
the weighted average  of the control unit times a positive constant is parallel to the sample path of the treated unit in the absence of treatment. 
%This modified synthetic control method (MSCM) contains the synthetic control method as a special case when the scale factor equals 1. 
Note that the MSCM approach contains the SCM as a special case which, in turn, include the DID method for a particular choice of weights. 
The MSCM  method performs well when the number of time periods (both pre and post) is large, but 
leads to larger estimation variance in the presence of a large number of control units. 
\citet{Hsiao/etc:2012} proposed a novel and flexible approach to estimate ATE which  requires neither the assumption of no sample selection effect 
nor that of the parallel paths for treatment and control units in the absence of treatment. 
Furthermore, their method, which we will refer to as the HCW method, yields consistent estimate of the ATE that is additionally robust to any nonlinear functional form. 
However, similar to the SCM, having a large number of control units can lead the HCW method to over-fit in-sample data 
and generate imprecise out-of-sample predictions. 

A major concern for all the above methods of estimating ATE from panel data is their extreme non-robustness against data contamination (e.g., outliers), 
as they are all averaging and/or least-squares techniques. Since outliers are not infrequent in real-life datasets,
these existing procedures lead to non-robust and incorrect inference about the ATE in the presence of data contamination.
When a treatment intervention is given to a particular unit, one may carefully collect its response during the follow-up (post-treatment) period to avoid any noise;
but it is not always feasible to protect the observations from contamination on all the control units (and also for the treated unit in pre-treatment time-points).
So, a robust procedure that can yield stable inference about the ATE from (possibly noisy) panel data is extremely necessary 
for correct policy planning and other insight generation in practical applications. 
The present paper solves this crucial issue by proposing a new robust estimation and subsequent inference procedures for the ATE 
under panel data set-up, leading to stable and highly efficient insights even under data contamination.

Among several existing approaches of robust parameter estimation, 
we consider the minimum distance methodology with the popular density power divergence (DPD) of \citet{Basu/etc:1998}.  
This minimum DPD estimation has become extremely popular in recent times due to its easy interpretation as a generalization of the maximum likelihood estimator (MLE)
and high asymptotic efficiency along with the desired robustness properties \cite{Basu/etc:2011}.
As a result, the minimum DPD estimator (MDPDE) has also been extended to several advanced data types and inference problems;
see, e.g., \cite{Ghosh/Basu:2013,Ghosh/Basu:2015,Ghosh/Basu:2016,Ghosh/Basu:2016b,Durio/Isaia:2011}.
For the sake of completeness, a brief review of the MDPDE is provided in Appendix \ref{APP:MDPDE}.
In this paper, we particularly use the MDPDE, as defined for the linear regression in \cite{Durio/Isaia:2011, Ghosh/Basu:2013}, 
along with the HCW factor model for panel data set-up to develop a new robust and efficient ATE estimator. 
The nice asymptotic properties of the MDPDE would translate to our proposed estimate of ATE establishing its consistency and asymptotic normality. 
Although the use of MDPDE in constructing the robust ATE estimator is somewhat straightforward, 
derivations of the asymptotic properties of the resulting ATE estimator are rather nontrivial 
which are the major contributions of the present paper along with their innovative applications in (robustly) studying treatment effects. 
Using the proposed ATE estimates, we also develop robust testing procedures for both one and two sample hypotheses involving ATE.
The claimed robustness of the proposed estimation and testing procedures, against data contamination in the pre-treatment periods,  
would be illustrated through appropriate Monte-Carlo simulation studies  and also theoretically via the influence function analysis of the ATE estimator.
An appropriate modification is also suggested to make the proposed ATE estimator robust against contamination in the post-treatment period's data, 
with empirical illustrations. 
Finally, our proposal is applied to examine the long-term effects of the 2004 Indian Ocean earthquake and tsunami on the economic growth (measured via GDP) 
of five countries, namely Indonesia, Sri Lanka, Thailand, India and Maldives, that had received the major instant shock.

The rest of the paper is structured as follows. 
In Section \ref{SEC:ATE-Est} we discuss the model setup and the proposed robust ATE estimator, 
while its asymptotic properties are derived in  Section \ref{SEC:Asymp}. 
Statistical tests based on the new ATE estimator are presented in Section \ref{SEC:Tests}.
Section \ref{SEC:SIMULATION} presents simulation studies to compare the efficacy of our proposed ATE estimators 
over the estimating ones and also to illustrate the finite sample performances of the associated robust tests. 
The (local) robustness of our ATE estimator is further studied theoretically in Section \ref{SEC:IF} through the influence function analysis,
where another modified version is discussed to gain robustness against data contamination in the post-treatment period. 
Section \ref{SEC:data} discusses the real data applications and some concluding remarks are presented in Section \ref{SEC:conclusion}.
For the brevity in presentation, some proofs are moved to Appendices \ref{APP:Proof_MDPDEs} and \ref{APP:Proof_Tests}.

\section{Robust Estimation of ATE from Panel Data}
\label{SEC:ATE-Est}

\subsection{Model Set-up}
\label{SEC:Model}

Suppose that we have a panel dataset consisting of observations on $N$ units over $T$ time periods,
and let $y_{it}$ denote the response of $i$-th unit at $t$-th time point for $i=1, \ldots, N$ and $t=1, \ldots, T$.
To identify the counterfactuals, let us also denote by $y_{it}^1$ and $y_{it}^0$ the outcome of unit  $i$ in period $t$ with and without treatment, respectively. 
Then the treatment effect of the $i^{th}$ unit at a post-treatment time-point  $t$ is defined as 
\begin{equation}\label{EQ:TE}
	\Delta_{it}=y_{it}^1 - y_{it}^0.
\end{equation}
However, for any $i$, $y_{it}^0$ and $y_{it}^1$ are not simultaneously observed at a time-point ($t$). 
Thus, the observed data actually have the form $y_{it}=d_{it}y_{it}^1+(1-d_{it})y_{it}^0$ for each $i$ and $t$, where
\begin{equation*}
	d_{it} = 
	\begin{cases}
		1, & \text{if
			the $i^{th}$ unit is under the treatment at time $t$ } \\
		0, &\text{otherwise}
	\end{cases}
\end{equation*}
The difficulty in estimating the counterfactual outcome $y_{it}^0$, when $d_{it}=1$, makes the estimation of ATE cumbersome.
For estimation of the counterfactual outcome of the treated unit in the post-treatment period, 
we consider the HCW factor model \cite{Hsiao/etc:2012} having several nice properties as described in the introduction (Section \ref{SEC:intro}).
It assumes the existence of $K$-dimensional unobservable (latent) factors $\boldsymbol{f}_t$ for every time point $t$ such that
\begin{equation}
	y_{it}^0 =a_i + \boldsymbol{b}_i'\boldsymbol{f}_t + u_{it}, ~~~~ i=1,...N,~~ t= 1,...T
	\label{EQ:Factor_model}
\end{equation}
where $a_i$ is a unit specific intercept, $\boldsymbol{b}_i$ is the $K$-dimensional factor loadings for $i$-th unit associated with the factor $\boldsymbol{f}_t$ 
and $u_{it}$ is a zero mean, weakly dependent and weakly stationary error term.
The above relation can also be written in a matrix form as
\begin{equation}\nonumber
	\boldsymbol{y}_t^0= \boldsymbol{a }+ \boldsymbol{B}\boldsymbol{f}_t + \boldsymbol{u}_t,~~~~~ t= 1,....,T
\end{equation}
where $\boldsymbol{y}_t^0= (y_{1t}^0,...y_{Nt}^0)'$,
% denotes the outcome variables at time $t$ without treatment, 
$ \boldsymbol{a} =(a_1,.....a_N)'$, $\boldsymbol{B}= [\boldsymbol{b}_1, ..., \boldsymbol{b}_N]'$ and $\boldsymbol{u}_t = (u_{1t},...u_{Nt})'$. 
%It has been suggested in \cite{Hsiao/etc:2012} that, when neither $N$ nor $T_1$ is large, the responses of the untreated (control) units 
%can be used in lieu of $\boldsymbol{f}_t$ to predict the counterfactual for the treated unit at time $t$.

%To illustrate the arguments more concretely, let us now 
Now, let us assume that one unit is exposed with the treatment at time $T_1$ $(0< T_1 < T-1)$;
without loss of generality, suppose that the first unit ($i=1$) is the treated one to avoid complex notation. 
Then, the responses of the first unit must satisfy the relations
\begin{eqnarray}
	y_{1t} = 
	\begin{cases}
		y_{1t}^0, & \text{ for } t=1,\ldots, T_1\\
		y_{1t}^1 = y_{1t}^0+\Delta_{1t} , & \text{ for } t=T_1+1,\ldots, T
	\end{cases}
	\label{2.5}
	%y_{1t}=y_{1t}^0 &=& a_1 + \boldsymbol{b}_1'\boldsymbol{f}_t + u_{it}, ~~~~~~~~~~~t=T_1+1,\ldots, T,
	%\nonumber\\
	%y_{1t}=y_{1t}^1 &=& y_{1t}^0+\Delta_{1t} = a_1 + \boldsymbol{b}_1'\boldsymbol{f}_t + \Delta_{1t}+ u_{it}, ~~~t=T_1+1,\ldots, T,
\end{eqnarray}
where the treatment effect $\Delta_{1t}$ is as defined in (\ref{EQ:TE}) for $i=1$.
Thus, once we obtain an estimate $\widehat{y_{1t}^0}$ of the counterfactual $y_{1t}^0$ for a post-treatment time point $t$,
the treatment effect at that time-point can be easily estimated as $\widehat{\Delta}_{1t} = y_{1t}-\widehat{y_{1t}^0}$;
these are finally averaged over all the post treatment periods to get an estimate of the ATE.  

However, the latent factors in (\ref{EQ:Factor_model}) are not observable and hence we cannot directly use it to estimate the counterfactual $y_{1t}^0$. 
An indirect way was proposed in \cite{Hsiao/etc:2012} which suggests to transform Equation (\ref{EQ:Factor_model}) as follows.  
For any $r$-vector $\boldsymbol{v}$, let us denote the $(r-1)$-vector obtained by removing the first element of $\boldsymbol{v}$ as $\widetilde{\boldsymbol{v}}$.
Now, choosing an appropriate vector $\boldsymbol{\gamma}\in\mathcal{N}(\boldsymbol{B}')$, the null space of $\boldsymbol{B}'$, 
with the first element of $\boldsymbol{\gamma}$ being unity (with suitable scaling), and pre-multiplying (\ref{EQ:Factor_model}) by ${\boldsymbol{\gamma}}'$,
we get
%$\widetilde{a}'\boldsymbol{B}=0$. Define $\widetilde{a}=(1,-\boldsymbol{\gamma})$. Then 
\begin{eqnarray}
	{\boldsymbol{\gamma}}'\boldsymbol{y}_t^0={\boldsymbol{\gamma}}'\boldsymbol{a}+\boldsymbol{\gamma}'\boldsymbol{B}\boldsymbol{f}_t+\boldsymbol{\gamma}'\boldsymbol{u}_t
	%\nonumber\\
	&\implies& y_{1t}^0+\widetilde{\boldsymbol{\gamma}}'\widetilde{\boldsymbol{y}_t^0}={\boldsymbol{\gamma}}'\boldsymbol{a}+\boldsymbol{\gamma}'\boldsymbol{u_t}
	\nonumber\\
	&\implies& y_{1t}^0=\delta_1 +\boldsymbol{\delta}^{\prime}\widetilde{\boldsymbol{y}_t^0}+\eta_{1t},
	\label{EQ:Factor_model_trans}
\end{eqnarray}
where $\delta_1={\boldsymbol{\gamma}}'\boldsymbol{a}$, $\boldsymbol{\delta}=\widetilde{\boldsymbol{\gamma}}'$, 
$\eta_{1t}=\boldsymbol{\gamma}'\boldsymbol{u_t}$
and $\widetilde{\boldsymbol{y}_t^0}=(y_{2t}^0,\cdots,y_{Nt}^0)$.
Note that, Equation (\ref{EQ:Factor_model_trans})  holds for every $t=1, \ldots, T$ and justify the suggestion of \cite{Hsiao/etc:2012}
that $\widetilde{\boldsymbol{y}_t^0}$ can be used in lieu of $\boldsymbol{f}_t $ to predict $y_{1t}^0$.
In particular, at the pre-treatment time periods we have $y_{it}^0=y_{it}$ so that (\ref{EQ:Factor_model_trans}) reduces to 
\begin{eqnarray}
	y_{1t}=\delta_1 +\boldsymbol{\delta}^{\prime}\widetilde{\boldsymbol{y}_t}+\eta_{1t}, ~~~~t=1, \ldots, T_1.
	\label{EQ:Factor_model_trans2}
\end{eqnarray}
The regression coefficients $\delta_1$ and $\boldsymbol{\delta}$ are then estimated by the least square method from (\ref{EQ:Factor_model_trans2}).
If these estimates are denoted by $\widehat{\delta_1}$ and $\widehat{\boldsymbol{\delta}}$,
then the counterfactual outcome $y_{1t}^0$, for the post-treatment period, can be estimated via  
$\widehat{y_{1t}^0}=\widehat{\delta_1}+\widehat{\boldsymbol{\delta}}'\widetilde{\boldsymbol{y}_t}$ for $t= T_1+1,...T$, 
since $y_{it}^0=y_{it}$ for all $t$ when $i\neq1$. 
%The treatment effect at period $t$, for $t \geq T_1+1$, is estimated by $\widehat{\Delta}_{1s}=y_{1s}-\widehat{y_{1s}^0}$, 

At this point, we should note that some assumptions are required for the derivation of (\ref{EQ:Factor_model_trans}) to go through 
as noted in \cite{Hsiao/etc:2012}; however, all the assumptions of \cite{Hsiao/etc:2012} were indeed not necessary as shown later in \cite{Li/Bell:2017b}. 
Accordingly, let us note down the following set of sufficient conditions.

%\noindent\textbf{Assumptions:}
\begin{enumerate}
	\item[(A1)] $ \boldsymbol{u}_t$ is weakly dependent with $ E(\boldsymbol{u}_t) =0$, $E(\boldsymbol{u}_t\boldsymbol{u}_t')=\boldsymbol{V}$, 
	an $N\times N$  diagonal matrix, $ E(\boldsymbol{u_tf_t}')=0$ for all $t$ and $ E(u_{jt}|d_{is})=0$ for all $j \neq i$. 
	
	\item[(A2)]  All rows of $\boldsymbol{B}$ are bounded (in $\ell_2$-norm) uniformly by a positive constant $m<\infty$  and  
	Rank$(\widetilde{\boldsymbol{B}})=K$, where $\widetilde{\boldsymbol{B}}$ is the $(N-1)\times K$ matrix obtained by removing the first row of $\boldsymbol{B}$.

	\item[(A3)]  For the post-treatment periods $t=T_1+1, \ldots, T$, $\eta_{1t}$ and $\Delta_{1t}$ are weakly dependent and weakly stationary processes 
	such that $\Delta_1=E(\Delta_{1t})$ for all $t$ and the central limit theorem applies to 
	$T_2^{-1/2} \sum_{t=T_1+1}^T \left(\Delta_{1t}-\Delta_1 +\eta_{1t}\right)$ having asymptotic mean zero and variance ${\Sigma}_2$. 
	
	% \item[A6.] For any fixed $K$ and $N$, there exists an $N\times1$ vector $\boldsymbol{\tilde{a}}^(\star)$ such that $\boldsymbol{\tilde{a}}^(\star)'B=0$, where $\mathcal{N}(\boldsymbol{B})$ is the null space of $\boldsymbol{B}$. At the neighbourhood of $\boldsymbol{\tilde{a}}^\star$, $T_1^{-1}\sum_{t=1}^{T_1}E(y_{1t}^0-\delta_1-\delta'\widetilde{y_t})^2$ has a unique minimum at $(\delta_{10},\boldsymbol{\delta_0}')'.$ 
	%\item[A7.] $E(u_{1t}^\star|\widetilde{\boldsymbol{y_t}})= c_1 + \boldsymbol{c'\widetilde{y_t}}$ where $u_{1t}^\star = \boldsymbol{\tilde{a}}'\boldsymbol{u_t}$ ( Assumption of linear conditional mean functional form).
\end{enumerate}

Note that only (A2) is needed for the calculations leading to (\ref{EQ:Factor_model_trans}).
Additionally, (A1) was used in \cite{Li/Bell:2017b} to show that the error term $\eta_{1t}$ and the coefficients $\delta_1$, $\boldsymbol{\delta}$ 
can be chosen (with some translation if required) so as to satisfy $E(\eta_{1t})=0$ and $E(\eta_{it}\widetilde{\boldsymbol{y}_t})=0$ for all $t$,
which ensure the consistency of the least-square based estimate of the ATE $\Delta_1$ along with (A3). 
Note also that, intuitively $\Delta_{1t}$ should be assumed to be independent of $\{\widetilde{\boldsymbol{y}_t}\}$ for any $t=1, \ldots, T$,
since the treatment effect should not depend on the control units. 

%
%We make the following assumptions:
%\begin{enumerate}
%	\item  [\hypertarget{Assumption 1}{i}] $\boldsymbol{\widetilde{y}_t}$ follow a $N-1$ dimensional multivariate normal distribution with mean vector $\boldsymbol{\mathbf{\mu}}$ and dispersion matrix $\boldsymbol{\Gamma}$ which is a diagonal matrix with diagonal elements $\gamma_i^2$, for all $i=2,..N$.
%\end{enumerate}
%
%For the post treatment periods, we add a treatment effect term to obtain
%\begin{equation}
%	y_{1s}=\delta_1+\boldsymbol{\widetilde{\delta}}'\boldsymbol{\widetilde{y_s}}+\Delta_{1s}^\alpha+v_{1s}, \hspace{2mm}  s=T_1+1,\cdots,T. 
%\end{equation} 

\subsection{A New Robust ATE Estimator using Density Power Divergence}

We now discuss our proposed approach for the robust estimation of ATE from panel data.
Let us consider the HCW factor model set-up and notation from the previous subsection so that 
the transformed Equations (\ref{EQ:Factor_model_trans}) and (\ref{EQ:Factor_model_trans2}) are valid  (requires Assumption (A2)). 
As mentioned before, we will use the popular minimum DPD estimation approach to get robust estimates of the parameters 
$\delta_1$, $\boldsymbol{\delta}$ and use them to derive the robust ATE estimator.  
In order to achieve greater efficiency, this approach of parametric robustness considers outliers defined with respect to a given model distribution 
that the majority of observations follow. Thus, in order to get such a parametric robust estimate of the regression coefficients from (\ref{EQ:Factor_model_trans2}),
we first consider a model density for the errors $\eta_{1t}$. So, instead of Assumption (A1), we will assume that these error terms,
after suitable adjustment through linear projection as in \cite{Li/Bell:2017b}, satisfy the following condition for a tuning parameter $\alpha\geq 0$. 

%\bigskip
%\noindent\textbf{Assumption (A4):}\\
\begin{enumerate}
	\item[(A4)] For $t=1, \ldots, T_1$, $\eta_{1t}$'s are independent and identically distributed (IID) each having density $\frac{1}{\sigma}f\left(\frac{\eta}{\sigma}\right)$,
	where $f$ is an univariate density with mean zero and variance one (e.g., standard normal or Laplace, etc.) 
	such that $M_f^{(\alpha)} = \int f^{1+\alpha}<\infty $ and $\sigma^2$ is the error variance (model parameter).
\end{enumerate}

Under Assumption (A4), it follows from Equation (\ref{EQ:Factor_model_trans2}) that
$y_{1t}$ has density $f_t(y_{1t}) = \frac{1}{\sigma}f\left(\frac{y_{1t}-\delta_1 -\boldsymbol{\delta}^{\prime}\widetilde{\boldsymbol{y}_t}}{\sigma}\right)$
for each $t=1, \ldots, T_1$ and they are independent given $\widetilde{\boldsymbol{y}_t}$.
Thus, as detailed in Appendix \ref{APP:MDPDE}, the MDPDE of the model parameters $\boldsymbol{\theta}=\left(\delta_1, \boldsymbol{\delta}, \sigma^2 \right)$,
at any given $\alpha>0$, can be obtained by minimizing the objective function (\ref{EQ:MDPDE_INH}) over $t=1, \ldots, T_1$ with $f_i \equiv f_t$, 
which simplifies to the form 
\begin{eqnarray}
	H_{T_1}^{(\alpha)}(\boldsymbol{\theta}) = \dfrac{1}{\sigma^\alpha}\left[M_f^{(\alpha)} - \left(1+\frac{1}{\alpha}\right)\frac{1}{T_1}\sum_{t=1}^{T_1} 
	f\left(\frac{y_{1t}-\delta_1 -\boldsymbol{\delta}'\widetilde{\boldsymbol{y}_t}}{\sigma}\right)^\alpha\right].
	\label{EQ:MDPDE_ATE}
\end{eqnarray}
In the particular case of normal error distribution, i.e., when $f$ is the standard normal density, the above MDPDE objective function can be further simplified as 
\begin{equation}
	H_{T_1}^{(\alpha)}(\boldsymbol{\theta})= 
	\frac{1}{(2\pi)^{\frac{\alpha}{2}}\sigma^\alpha\sqrt{1+\alpha}}-\frac{(1+\alpha)}{\alpha(2\pi)^{\alpha/2}\sigma^\alpha}\frac{1}{T_1}\sum_{t=1}^{T_1}
	e^{-\frac{\alpha}{2\sigma^2}\left(y_{1t}-\delta_1 -\boldsymbol{\delta}'\widetilde{\boldsymbol{y}_t}\right)^2},
	\label{EQ:MDPDE_ATE0}
\end{equation} 
which we need to minimize with respect to $\boldsymbol{\theta}$ to obtain the MDPDEs of the regression coefficients and the error variance. 
Note that, unlike the least squared based methods, the MDPDE objective function $H_{T_1}^{(\alpha)}(\boldsymbol{\theta})$ in (\ref{EQ:MDPDE_ATE}),
or its simpler version in (\ref{EQ:MDPDE_ATE0}), does not have an explicit minimizer. 
So, we need to compute the MDPDE by numerically minimizing $H_{T_1}^{(\alpha)}(\boldsymbol{\theta})$ via an appropriate optimization technique. 

Equivalently, the MDPDEs can also be obtained by numerically solving the corresponding estimating equations obtained 
by equating the derivative of $H_{T_1}^{(\alpha)}(\boldsymbol{\theta})$, with respect to $\boldsymbol{\theta}$, to zero. 
For the special case of normal error density, by differentiating $H_{T_1}^{(\alpha)}(\boldsymbol{\theta})$ in (\ref{EQ:MDPDE_ATE0}) 
with respect to $\delta_1$, $\delta_j$, $j=2,3,\cdots N$ and $\sigma$, we get the corresponding MDPDE estimating equations as given by 

\begin{eqnarray}
	\sum_{t=1}^{T_1}\left(y_{1t}-\delta_1 -\boldsymbol{\delta}'\widetilde{\boldsymbol{y}_t}\right)
	e^{-\frac{\alpha}{2\sigma^2}\left(y_{1t}-\delta_1 -\boldsymbol{\delta}'\widetilde{\boldsymbol{y}_t}\right)^2}&=&0
	\label{2.12} \\
	\sum_{t=1}^{T_1}y_{jt}\left(y_{1t}-\delta_1 -\boldsymbol{\delta}'\widetilde{\boldsymbol{y}_t}\right)
	e^{-\frac{\alpha}{2\sigma^2}\left(y_{1t}-\delta_1 -\boldsymbol{\delta}'\widetilde{\boldsymbol{y}_t}\right)^2} &=& 0, ~~~~~~~j=2,3,\cdots N
	\label{2.13} \\ 
	\sum_{i=1}^{T_1}\left[1-\frac{\left(y_{1t}-\delta_1 -\boldsymbol{\delta}'\widetilde{\boldsymbol{y}_t}\right)^2}{\sigma^2}\right]
	e^{-\frac{\alpha}{2\sigma^2}\left(y_{1t}-\delta_1 -\boldsymbol{\delta}'\widetilde{\boldsymbol{y}_t}\right)^2} &=& \frac{\alpha}{(1+\alpha)^{\frac{3}{2}}}.
	\label{2.14}
\end{eqnarray}
For the general error distribution $f$, assuming $f$ to be differentiable and letting $u=f'/f$,  
the general set of MDPDE estimating equations can be derived as 
%by considering the derivative of (\ref{EQ:MDPDE_ATE}) with respect to $\boldsymbol{\theta}$ and are given by 
\begin{eqnarray}
	\sum_{i=1}^{n}f\left(\frac{y_{1t}-\delta_1-\boldsymbol{\delta}'\widetilde{\boldsymbol{y}_t}}{\sigma}\right)^\alpha u\left(\frac{y_{1t}-\delta_1-\boldsymbol{\delta}'\widetilde{\boldsymbol{y}_t}}{\sigma}\right)&=&0,\\
	\sum_{i=1}^{n}f\left(\frac{y_{1t}-\delta_1-\boldsymbol{\delta}'\widetilde{\boldsymbol{y}_t}}{\sigma}\right)^\alpha u\left(\frac{y_{1t}-\delta_1-\boldsymbol{\delta}'\widetilde{\boldsymbol{y}_t}}{\sigma}\right)\widetilde{\boldsymbol{y}_t}&=&0,\\
\frac{1}{T_1}\sum_{t=1}^{T_1}\left[1+ \left(\frac{y_{1t}-\delta_1-\boldsymbol{\delta}'\widetilde{\boldsymbol{y}_t}}{\sigma}\right) 
u\left(\frac{y_{1t}-\delta_1-\boldsymbol{\delta}'\widetilde{\boldsymbol{y}_t}}{\sigma}\right)\right]
f\left(\frac{y_{1t}-\delta_1-\boldsymbol{\delta}'\widetilde{\boldsymbol{y}_t}}{\sigma}\right)^\alpha &=& \frac{\alpha M_f^{(\alpha)}}{1+\alpha}.
~~~~~
\end{eqnarray}
%where we denote $u=f'/f$ and $M_{f,i,j}^{(\alpha)} = \int u^i (u')^j f^{1+\alpha}$ for any $i,j=0, 1, 2, \ldots,$ and any $\alpha\geq 0$. 

Let us now denote the MDPDE of $\boldsymbol{\theta}=\left(\delta_1, \boldsymbol{\delta}, \sigma^2 \right)$, obtained with tuning parameter $\alpha> 0$,
as $\widehat{\boldsymbol{\theta}}_\alpha=\left(\widehat{\delta}_{1,\alpha}, \widehat{\boldsymbol{\delta}}_\alpha, \widehat{\sigma}_\alpha^2 \right)$.
Using them, we can (robustly) estimate the  counterfactual outcome $y_{1t}^0$ at any post-treatment time-period as
\begin{equation}\nonumber
	\widehat{y}_{1t}^0 = \widehat{\delta}_{1,\alpha} + \widehat{\boldsymbol{\delta}}_\alpha'\widetilde{\boldsymbol{y}_t}, ~~~t=T_1+1, \ldots, T,
\end{equation}
and the corresponding  treatment effects are estimated by
\begin{equation}
	\widehat{\Delta}_{1t}^{(\alpha)} = y_{1t} - \widehat{y}_{1t}^0, ~~ t=T_1+1, \ldots, T.\nonumber
\end{equation} 
Finally, a robust estimate of the ATE $\Delta_1$ is given by 
\begin{equation}\label{EQ:Mean-MDPDE}
	\widehat{\Delta}_1^{(\alpha)} = \frac{1}{T_2}\sum_{t=T_1+1}^{T}\widehat{\Delta}_{1t}^{(\alpha)}.
\end{equation}
Due to the use of the MDPDE of the regression coefficients $(\delta_1, \boldsymbol{\delta})$ and the averaging at the final step, 
we will refer to the estimator $\widehat{\Delta}_1^{(\alpha)} $ as the Mean-MDPDE of the ATE $\Delta_1$ with tuning parameter $\alpha>0$.

\begin{remark}[Special Case: $\alpha\rightarrow 0$]
Since the MDPDE at $\alpha\rightarrow0$ coincides with the MLE of the regression parameter 
(which is also the least square estimate for regression coefficients under normal error model), 
the proposed Mean-MDPDE of the ATE becomes the classical HCW estimator \cite{Hsiao/etc:2012} as $\alpha\rightarrow0$.
Thus, for any $\alpha>0$, the Mean-MDPDE of the ATE indeed provides a robust generalization of the HCW estimator. 
\end{remark}

\section{Asymptotic Properties of the Proposed Estimators}
\label{SEC:Asymp}

\subsection{Properties of the MDPDE of the Regression Parameters}
\label{SEC:Asymp_MDPDE}

%In this section, we discuss the large sample behavior of the MDPDE of the  regression coefficients from \cite{Ghosh/Basu:2013},
%and use them to show the consistency and asymptotic normality of the proposed robust  ATE estimator.
Let us first discuss the  large sample behavior of the MDPDE of the  regression coefficients and the error variance following \cite{Ghosh/Basu:2013}.
In this context, we need some additional assumptions as follows.

\begin{enumerate}
	\item[(A5)] Let us denote $\boldsymbol{x}_t=(1,\widetilde{\boldsymbol{y}_t}')'=(x_{t1}, \ldots, x_{tN})'$ for each $t=1, \ldots, T$ and define 
	$\boldsymbol{X}_0' =[\boldsymbol{x}_1, \cdots, \boldsymbol{x}_{T_1}]$ and  $\boldsymbol{X}_1' =[\boldsymbol{x}_{T_1+1}, \cdots, \boldsymbol{x}_{T_2}]$.
	Then, with probability tending to one, the following results hold: 
	\begin{eqnarray}
		& \sup\limits_{T_1>1} \max\limits_{1\leq t \leq T_1}	 |x_{tj}| = O(1), ~~\sup\limits_{T_1>1} \max\limits_{1\leq t\leq T_1} |{x_{tj}}{x_{tk}}| = O(1),~~
		\frac{1}{T_1}\sum_{t=1}^{T_1} |{x_{tj}}{x_{tk}}{x_{tl}}|=O(1), 
		\nonumber\\
		&~~~~~~~~~~~~~~~~~~\mbox{ for all }j, k, l = 1, \ldots, N,~~~
		\nonumber\\ 
		&\inf\limits_{T_1>1}\left[\mbox{min eigenvalue of } T_1^{-1}(\boldsymbol{X}_0'\boldsymbol{X}_0)\right] > 0, ~~~ \mbox{ and }~~
		\max\limits_{1\leq t \leq T_1}[\boldsymbol{x}_t'(\boldsymbol{X}_0'\boldsymbol{X}_0)^{-1}\boldsymbol{x}_t]=O(T_1^{-1}).
		\nonumber
	\end{eqnarray}
\end{enumerate}

%Consider the  normal linear regression model: 
%\begin{equation}
%    y_{1t}=\delta_1+ \boldsymbol{\delta'\widetilde{y_t}}+\eta_{1t}=\boldsymbol{\beta'{x_t}}+\eta_{1t}, i=1(1)n
%\end{equation} where $\boldsymbol{\beta}=(\delta_1,\boldsymbol{\delta})'$,  and the errors $\eta_{1t}$'s are iid normal variables with mean zero and variance $\sigma^2$. Similar to Ghosh \& Basu (2013) \hyperlink{ref6}{[6]}, we assume that the true data generating density belongs to the model family and the given values of the independent variables($\boldsymbol{x_t}$) satisfy the following assumptions:-\\ 

It is easy to see that Assumption (A5) holds if $\{\boldsymbol{x}_t\}$ is a weakly dependent and weakly stationary process 
with $\boldsymbol{\Sigma}_x = \displaystyle {\mbox{plim}}_{T_1\rightarrow\infty }T_1^{-1}(\boldsymbol{X}_0'\boldsymbol{X}_0)$ being invertible. 
Then, the asymptotic properties of the MDPDEs 
$\widehat{\boldsymbol{\beta}}_\alpha=\left(\widehat{\delta}_{1,\alpha}, \widehat{\boldsymbol{\delta}}_\alpha' \right)$ and  $\widehat{\sigma}_\alpha^2$
of the regression parameters  $\boldsymbol{\beta}=\left(\delta_1, \widetilde{\delta}' \right)$ and $\sigma^2$, respectively, 
can be obtained by an application of the general theory from \cite{Ghosh/Basu:2013}. 
Specifically, under Assumptions (A4)--(A5), we can show the following result for any $\alpha\geq 0$;
see Appendix \ref{APP:Proof_MDPDEs} for proof.

\begin{theorem}[Asymptotics of the MDPDEs under linear regression with a general error density]
	\label{PROP:Asymp_MDPDE}
Under Assumptions (A4)--(A5) with general error density $f$, 
there exists a consistent sequence of roots (MDPDEs) $\widehat{\boldsymbol{\beta}}_\alpha$ and $\widehat{\sigma}_\alpha^2$ 
to the minimum DPD estimating equations \eqref{2.12}-\eqref{2.14}, which are $\sqrt{T_1}$-consistent for the true parameter value $\boldsymbol{\beta}$ and $\sigma^2$,
respectively, and  
$$
\sqrt{T_1}\left(\begin{array}{c}
		\widehat{\boldsymbol{\beta}}_\alpha -\boldsymbol{\beta}\\
		\widehat{\sigma}_\alpha^2-\sigma^2
	\end{array}\right)
\mathop{\rightarrow}^{\mathcal{D}} 
\mathcal{N}_{N+1}\left(\boldsymbol{0}, \sigma^2\widetilde{\boldsymbol{\Psi}}_\alpha^{-1}\widetilde{\boldsymbol{\Omega}}_\alpha\widetilde{\boldsymbol{\Psi}}_\alpha^{-1}
\right),
~~~~~~~~
\mbox{as } T_1 \rightarrow \infty, 
$$
where 
%$\boldsymbol{V}=\widetilde{\boldsymbol{\Psi}}_\alpha^{-1}\widetilde{\boldsymbol{\Omega}}_\alpha\widetilde{\boldsymbol{\Psi}}_\alpha^{-1}$,  
\begin{align}
\widetilde{\boldsymbol{\Psi}}_\alpha 
= \left(\begin{array}{cc}
		\zeta_{11,\alpha}\boldsymbol{\Sigma}_x & \frac{\zeta_{12,\alpha} }{\sigma}\boldsymbol{\mu}_x\\
		\frac{\zeta_{12,\alpha} }{\sigma}\boldsymbol{\mu}_x & \frac{\zeta_{22,\alpha}}{\sigma^2}
\end{array}\right),
~~~~~
\widetilde{\boldsymbol{\Omega}}_\alpha = \left(\begin{array}{cc}
		[\zeta_{11,2\alpha}-\phi_{1,\alpha}^2] \boldsymbol{\Sigma}_x & \frac{[\zeta_{12,2\alpha}-\phi_{1,\alpha}\phi_{2,\alpha}]}{\sigma}\boldsymbol{\mu}_x\\
		\frac{[\zeta_{12,2\alpha}-\phi_{1,\alpha}\phi_{2,\alpha}] }{\sigma}\boldsymbol{\mu}_x & \frac{[\zeta_{22,2\alpha}-\phi_{2,\alpha}^2]}{\sigma^2}
	\end{array}\right), \nonumber
\end{align}
with 
%$\boldsymbol{\Sigma}_x = \displaystyle{\mbox{plim}}\lim\limits_{T_1\rightarrow\infty }T_1^{-1}(\boldsymbol{X}_0'\boldsymbol{X}_0)$,   
$\boldsymbol{\mu}_x= \displaystyle{\mbox{plim}}\lim\limits_{T_1\rightarrow\infty }T_1^{-1}(\boldsymbol{X}_0'\boldsymbol{1})$,
$\boldsymbol{1}$ denotes a vector of appropriate length with all entries being one 
and
$$
\zeta_{11, \alpha}= M_{f,0,2}^{(\alpha)},
~~
\zeta_{12,\alpha}=\left[M_{f,0,1}^{(\alpha)}+M_{f,1,2}^{(\alpha)}\right]/2, 
~~
\zeta_{22,\alpha}=\left[M_{f,2,2}^{(\alpha)}+ 2 M_{f,1,1}^{(\alpha)}+M_{f,0,0}^{(\alpha)}\right]/4,
$$
$$
\phi_{1,\alpha}= - M_{f,0,1}^{(\alpha)},  
~~~~\phi_{2,\alpha}= - \left[M_{f,0,0}^{(\alpha)}+M_{f,1,1}^{(\alpha)}\right]/2,
$$
with $M_{f,i,j}^{(\alpha)}=\int s^i u(s)^j f(s)^{1+\alpha}ds$ for any $i,j=0,1,2$. 
\end{theorem}

%\bigskip
\begin{corollary}\label{CORR:Asymp_MDPDE}
Under the assumptions of Theorem \ref{PROP:Asymp_MDPDE}, 
additionally if the error density $f$ satisfies $M_{f,0,1}^{(\alpha)}=0=M_{f,1,2}^{(\alpha)}$ for every $\alpha>0$
then the MDPDEs $\widehat{\boldsymbol{\beta}}_\alpha$ and $\widehat{\sigma}_\alpha^2$ are asymptotically independent and 
their asymptotic distributions are explicitly given by 
$$
\sqrt{T_1}\left(\widehat{\boldsymbol{\beta}}_\alpha -\boldsymbol{\beta}\right) \mathop{\rightarrow}^{\mathcal{D}} 
\mathcal{N}_{N}\left(\boldsymbol{0}, \sigma^2v_\beta(\alpha) \boldsymbol{\Sigma}_x^{-1}\right),
~~~~~~~~
\mbox{as } T_1 \rightarrow \infty, 
$$
and
$$
\sqrt{T_1}\left(\widehat{\sigma}_\alpha^2-\sigma^2\right) \mathop{\rightarrow}^{\mathcal{D}} 
\mathcal{N}\left(\boldsymbol{0}, \sigma^4 v_\sigma(\alpha)\right),
~~~~~~~~
\mbox{as } T_1 \rightarrow \infty, 
$$
where $v_\beta(\alpha) = \left[\zeta_{11,2\alpha}-\phi_{1,\alpha}^2\right]\zeta_{11,\alpha}^{-2}$ 
and $v_\sigma(\alpha) = \left[\zeta_{22,2\alpha}-\phi_{2,\alpha}^2\right]\zeta_{22,\alpha}^{-2}$. 
\end{corollary}

\medskip
In particular, when the error density $f$ is standard normal, the conditions of Corollary \ref{CORR:Asymp_MDPDE} hold true 
and the asymptotic variances of the MDPDEs $\widehat{\boldsymbol{\beta}}_\alpha$ and $\widehat{\sigma}_\alpha^2$ can be further simplified with
\begin{eqnarray}
	v_\beta(\alpha) &=& \left(1+\frac{\alpha^2}{1+2\alpha}\right)^{3/2}, 
	\nonumber\\
	v_\sigma(\alpha)&=&\frac{4}{(\alpha^2+2)^2}\left[2(1+2\alpha^2)\left(1+\frac{\alpha^2}{1+2\alpha}\right)^{5/2}-\alpha^2(1+\alpha)^2\right]. 
	\nonumber
\end{eqnarray}

These asymptotic distributional results of the MDPDEs will be used, in the following subsection,  
to derive the consistency and asymptotic distribution of the proposed Mean-MDPDE estimator $\widehat{\Delta}_1^{(\alpha)}$ of the ATE $\Delta_{1}$ 
as $T_1, T_2 \rightarrow\infty$. 
For simplicity, throughout the rest of the paper, we will assume that the conditions of Corollary \ref{CORR:Asymp_MDPDE} hold
so that the asymptotic distributions of $\widehat{\boldsymbol{\beta}}_\alpha$ and $\widehat{\sigma}_\alpha^2$ are independent;
all the results, however, can easily be extended for the general case using Theorem \ref{PROP:Asymp_MDPDE} with messier formulas and notations.

\subsection{Asymptotics for the proposed ATE Estimator}
\label{SEC:Asymp_ATE}

In order to derive the asymptotic properties of the ATE estimator $\widehat{\Delta}_1^{(\alpha)} $, 
let us first note its decomposition as given by 
\begin{eqnarray}
	\widehat{\Delta}_1^{(\alpha)}  - \Delta_{1} &=&  \frac{1}{T_2}\sum_{t=T_1+1}^{T}\left(y_{1t} - \widehat{y}_{1t}^0\right) - \Delta_{1}
	\nonumber\\
	&=& \frac{1}{T_2}\sum_{t=T_1+1}^{T}\left( \delta_1 +\boldsymbol{\delta}^{\prime}\widetilde{\boldsymbol{y}_t}+\eta_{1t} + \Delta_{1t}- \widehat{\delta}_{1,\alpha} - \widehat{\boldsymbol{\delta}}_\alpha'\widetilde{\boldsymbol{y}_t} \right) - \Delta_1 .
	\nonumber
\end{eqnarray}
After some algebra, we get 
\begin{eqnarray}
	\widehat{\Delta}_1^{(\alpha)}  - \Delta_{1} 
	= -  \left(\widehat{\boldsymbol{\beta}}_\alpha - \boldsymbol{\beta}\right)' \left[\frac{1}{T_2}\sum_{t=T_1+1}^{T}\boldsymbol{x}_t\right] 
	+ \frac{1}{T_2}\sum_{t=T_1+1}^{T}\left(\Delta_{1t} - \Delta_1 + \eta_{1t} \right).
	\label{EQ:Main}
\end{eqnarray}
Note that, by Theorem \ref{PROP:Asymp_MDPDE}, $\left(\widehat{\boldsymbol{\beta}}_\alpha - \boldsymbol{\beta}\right) = O_p(T_1^{-1/2})$ as $T_1\rightarrow\infty$.
Further, via the laws of large numbers, we get  
\begin{eqnarray}
	\left[\frac{1}{T_2}\sum_{t=T_1+1}^{T}\boldsymbol{x}_t\right] \mathop{\rightarrow}^{\mathcal{P}} E[\boldsymbol{x}_t]<\infty, 
	~~~\mbox{ as }T_2\rightarrow\infty.
	\label{EQ:eq1}
\end{eqnarray}
Therefore, the first term in (\ref{EQ:Main}) is $O_p(T_1^{-1/2})$ as $T_1, T_2\rightarrow\infty$. 
Additionally, due to Assumption (A3), we have $T_2^{-1/2}\sum_{t=T_1+1}^{T}\left(\Delta_{1t} - \Delta_1 + \eta_{1t} \right)=O_p(1)$. 
Hence, combining all these, we get the consistency of the Mean-MDPDE $\widehat{\Delta}_1^{(\alpha)}$ as presented in the following theorem. 

\begin{theorem}\label{THM:Consist_ATE}
	For any given $\alpha\geq 0$, if Assumptions (A1)--(A5) hold, then we have 
	$$
	\widehat{\Delta}_1^{(\alpha)} - \Delta_1 = O_p\left(T_1^{-1/2}+T_2^{-1/2}\right), ~~~~\mbox{ as } T_1, T_2 \rightarrow\infty.
	$$
\end{theorem}

Next, in order to derive the asymptotic distribution of the ATE estimator $\widehat{\Delta}_1^{(\alpha)}$, 
we need appropriate assumption to make the two terms in (\ref{EQ:Main}) asymptotically independent. 
In this respect, we consider the following general assumption.

\begin{enumerate}
	\item[(A6)] The MDPDE $\widehat{\boldsymbol{\beta}}_\alpha$ is asymptotically independent of $\boldsymbol{x}_t$ and $\eta_{1t}$ for $t=T_1+1, \ldots, T_2$. 
\end{enumerate}

Note that Assumption (A6) can be shown to hold in several practical scenarios under suitable mixing conditions
for the processes $\{\boldsymbol{x}_t, \eta_{1t}\}$ for the whole time-period $t=1, \ldots, T$. 
Such a mixing condition is also used in \cite{Li/Bell:2017b} while deriving the asymptotic normality of their ATE estimator. 
Particularly, (A6) holds if the variables from the pre-treatment period are independent of their values in the post-treatment period, 
e.g., if $(\boldsymbol{x}_t, \eta_{1t})$s are independently distributed over $t$. 
With this additional assumption, we can now derive the asymptotic distribution of  $\widehat{\Delta}_1^{(\alpha)}$
by computing the asymptotic variances of the two terms in (\ref{EQ:Main}) individually, which is presented in the following theorem.

\begin{theorem}\label{THM:Asymp_ATE}
	For any given $\alpha\geq 0$, if Assumptions (A1)--(A6) hold and $\omega:=\lim\limits_{T_1, T_2\rightarrow\infty }T_2/T_1 <\infty$, 
	then we have 
	$$
	\sqrt{T_2}\left(\widehat{\Delta}_1^{(\alpha)} - \Delta_1 \right) \mathop{\rightarrow}^{\mathcal{D}} \mathcal{N}(0,\Sigma(\alpha)),~~~~\mbox{ as } T_1, T_2 \rightarrow\infty. 
	$$  
	where 
	$$
	\Sigma(\alpha) =  v_\beta(\alpha) ~ \omega \sigma^2E[\boldsymbol{x}_t]'\boldsymbol{\Sigma}_x^{-1}E[\boldsymbol{x}_t] + {\Sigma}_2.
	$$
\end{theorem} 
\textbf{Proof:}
Let us start from Equation (\ref{EQ:Main}) to deduce
$\sqrt{T_2}\left(\widehat{\Delta}_1^{(\alpha)}  - \Delta_{1} \right) = S_1 + S_2$,
%	\label{EQ:eq2}
%\end{eqnarray}
where
\begin{eqnarray}
	S_1	&=& - \sqrt{\frac{T_2}{T_1}}  \sqrt{T_1}\left(\widehat{\boldsymbol{\beta}}_\alpha - \boldsymbol{\beta}\right)'
	\left[\frac{1}{T_2}\sum_{t=T_1+1}^{T}\boldsymbol{x}_t\right], 
	\nonumber\\
	S_2 &=& \frac{1}{\sqrt{T_2}}\sum_{t=T_1+1}^{T}\left(\Delta_{1t} - \Delta_1 + \eta_{1t} \right).
\end{eqnarray}

By Assumption (A3), we directly get that $S$ is asymptotically normal with mean zero and variance ${\Sigma}_2$ as $T_2 \rightarrow\infty$.  . 
Therefore, the theorem will follow in view of  Assumption (A6) if we can show that $S_1$ is asymptotically normal with mean zero and variance 
$\boldsymbol{\Sigma}_1 =  \omega v_\beta(\alpha) \sigma^2 E[\boldsymbol{x}_t]'\boldsymbol{\Sigma}_xE[\boldsymbol{x}_t] $ as $T_1, T_2 \rightarrow\infty$.

But, by Theorem \ref{PROP:Asymp_MDPDE}, we have that $\sqrt{T_1}\left(\widehat{\boldsymbol{\beta}}_\alpha - \boldsymbol{\beta}\right)$ 
is asymptotically normal with mean zero and
variance $v_\beta(\alpha)\sigma^2\boldsymbol{\Sigma}_x^{-1}$ and that $\left[\frac{1}{T_2}\sum_{t=T_1+1}^{T}\boldsymbol{x}_t\right]$ converges in probability to $E[\boldsymbol{x}_t]$,
viz. Eq. (\ref{EQ:eq1}), as $T_1, T_2 \rightarrow\infty$.  Combining them via Slutsky's theorem and Assumption (A6), we get the asymptotic normality 
of $S_1$ with the asymptotic mean zero and variance 
$\left(\sqrt{\omega}  E[\boldsymbol{x}_t]\right)'[v_\beta(\alpha)\boldsymbol{\Sigma}_x]\left(\sqrt{\omega}  E[\boldsymbol{x}_t]\right) = \boldsymbol{\Sigma}_1$.
This completes the proof.
\hfill{$\square$}

\begin{figure}[!b]
	\centering
	\includegraphics[width=0.45\textwidth]{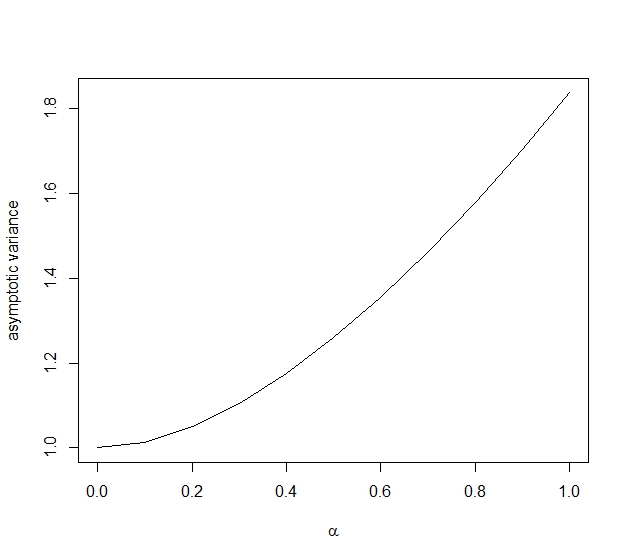}
	\caption{Plot of $v_\beta(\alpha)$ in the asymptotic variance of the Mean-MDPDE of the ATE, over $\alpha$, for normally distributed errors}
	\label{FIG:Asymp_Var}
\end{figure}

\bigskip
It is important to note that the asymptotic variance of the Mean-MDPDE of ATE depends on the tuning parameter $\alpha$ only through the term $v_\beta(\alpha)$.
For the normal error distribution, $v_\beta(0)=1$ and hence the asymptotic variance $\Sigma(\alpha=0) $ coincides with the asymptotic variance of the HCW estimator
as given in Theorem 3.2 of \cite{Li/Bell:2017b}. As $\alpha>0$ increases, the values of  $v_\beta(\alpha)$ increases leading 
to an increase in the asymptotic variance of the proposed ATE estimator, a price in order to achieve robustness under data contamination. 
However, by examining the values of $v_\beta(\alpha)$, e.g., in Figure \ref{FIG:Asymp_Var}, 
it can be seen that the loss in efficiency (in terms of the inverse of the asymptotic variance) is not quite significant 
at small values of the tuning parameter $\alpha>0$. 
This fact that $\alpha$ provides a trade-off between robustness and efficiency of the proposed ATE estimator is consistent 
with the literature of MDPDE \cite{Basu/etc:2011} and, in general, the notion of parametric robustness \cite{Hampel/etc:1986}.

\subsection{Consistent Estimation of the Standard Error}
\label{SEC:Asymp_VarEst}

Once we have an estimate of the ATE, we need to compute its standard error in order to have an idea about the credibility of our ATE estimate. 
For this purpose, we can use Theorem \ref{THM:Asymp_ATE} to get an asymptotic standard error for the proposed Mean-MDPDE of the ATE 
by consistently estimating the asymptotic variance $\Sigma(\alpha)$. If $\widehat{\Sigma}(\alpha)$ denote its consistent estimator, 
then the estimated asymptotic standard error of the ATE estimator $\widehat{\Delta}_1^{(\alpha)}$ is given by $\sqrt{\widehat{\Sigma}(\alpha)/T_2}$.

Now, in order to get the consistent estimate of $\Sigma(\alpha)$, we note that its second component ${\Sigma}_2$ is defined in Assumption (A3) 
in exactly the same way as in \cite{Li/Bell:2017b}. Hence, following the arguments given in \cite{Li/Bell:2017b}, a consistent estimator of ${\Sigma}_2$ is given by 
\begin{eqnarray}
	\widehat{{\Sigma}_2} = \frac{1}{T_2} \sum_{t=T_1+1}^T \sum_{s=T_1+1, |t-s|\leq l}^T \left(\widehat{\Delta}_{1t}^{(\alpha)} - \widehat{\Delta}_1^{(\alpha)}\right)
	\left(\widehat{\Delta}_{1s}^{(\alpha)} - \widehat{\Delta}_1^{(\alpha)}\right),
	\label{EQ:Sigma2_1}
\end{eqnarray} 
where $l$ is chosen in such a way that $l\rightarrow\infty$ and $l/T_2 \rightarrow 0$ as $T_2 \rightarrow\infty$
(e.g., $l=O(T_2^{1/4})$, see also \cite{Newey/West:1987,White:2001}). 
However, if $\Delta_{1t}$ and $\eta_{1t}$ are serially uncorrelated, then an alternative simpler consistent estimate of ${\Sigma}_2$ can be obtained as \citep{Li/Bell:2017b}
\begin{eqnarray}
	\widehat{{\Sigma}_2} = \frac{1}{T_2} \sum_{t=T_1+1}^T  \left(\widehat{\Delta}_{1t}^{(\alpha)} - \widehat{\Delta}_1^{(\alpha)}\right)^2.
	\label{EQ:Sigma2_2}
\end{eqnarray} 

Next, in order to consistently estimate the first component of $\Sigma$, we note that the term $\boldsymbol{\Sigma}_x$ can be consistently estimated by 
$T_1^{-1}(\boldsymbol{X}_0'\boldsymbol{X}_0)$. Similarly, a consistent estimate of  $E[\boldsymbol{x}_t]$ is given by $\left[\frac{1}{T_2}\sum_{t=T_1+1}^{T}\boldsymbol{x}_t\right]$. 
On the other hand, the MDPDE $\widehat{\sigma}_\alpha^2$ is consistent for $\sigma^2$ in view of Theorem \ref{PROP:Asymp_MDPDE}.
So, estimating $\omega$ by $T_2/T_1$, a consistent estimate of the first term of $\Sigma(\alpha)$ 
is then given by 
$$
v_\beta(\alpha) ~ \frac{T_2}{T_1} ~\widehat{\sigma}_\alpha^2 \left[\frac{1}{T_2}\sum_{t=T_1+1}^{T}\boldsymbol{x}_t\right]'
T_1(\boldsymbol{X}_0'\boldsymbol{X}_0)^{-1}\left[\frac{1}{T_2}\sum_{t=T_1+1}^{T}\boldsymbol{x}_t\right].
$$
Upon simplification, the final consistent estimate of the asymptotic variance $\Sigma(\alpha)$ is given by 
\begin{eqnarray}
	\widehat{\Sigma}(\alpha) = \frac{v_\beta(\alpha) \widehat{\sigma}_\alpha^2}{T_2} \left[\sum_{t=T_1+1}^{T}\boldsymbol{x}_t\right]'
	\left[\sum_{t=1}^{T_1}\boldsymbol{x}_t\boldsymbol{x}_t'\right]^{-1}\left[\sum_{t=T_1+1}^{T}\boldsymbol{x}_t\right] + \widehat{{\Sigma}_2},
	\label{EQ:Var_Est}
\end{eqnarray}
where $\widehat{{\Sigma}_2}$ can be taken as in (\ref{EQ:Sigma2_1}) or (\ref{EQ:Sigma2_2}) depending on our assumption.

%In practice, $\Sigma^2$ will be unknown. Thus, we need to find a consistent estimator of $\Sigma^2$. A consistent estimator of $\Sigma^2$ is given by 
%\begin{equation}\label{4.2}
%    \widehat{\Sigma^2}= \widehat{v_\alpha^b}\bigg[\widehat{E}(\sum_{s=T_1+1}^{T_2}\widetilde{x_s})'\widehat{E} (\sum_{s=1}^{T_1}\widetilde{x_s}'\widetilde{x_s})^{-1}\widehat{E}(\sum_{s=T_1+1}^{T_2}\widetilde{x_s})\bigg]+\widehat{\tau^2}+\widehat{\sigma^2}
%\end{equation}
%
%where $\widehat{E}(\sum_{s=T_1+1}^{T}\widetilde{x_s})=\widehat{E}(\sum_{s=T_1+1}^{T} (1,\widetilde{y_s}))=(T_2,\sum_{s=T_1+1}^{T}\widetilde{y_s})$, $ \widehat{E}(\sum_{s=1}^{T_1}\widetilde{x_s}'\widetilde{x_s})=\sum_{s=1}^{T_1}\widetilde{x_s}'\widetilde{x_s}$ , $\widehat{v_\alpha^b}=\widehat{\sigma}^2(1+\frac{\alpha^2}{1+2\alpha})^{3/2}$ and 
%$\widehat{\tau}^2=\frac{1}{T_2}\sum_{s=T_1+1}^{T_2}(\widehat{\Delta}_{1s}-\bar{\widehat{\Delta}}_{1s})^2$ and $\widehat{\sigma}^2 $ is obtained from the estimating equations \eqref{2.12}-\eqref{2.14} .

%\newpage
\section{Application in Robust Testing of Hypotheses on ATE}
\label{SEC:Tests}

%Testing of statistical hypothesis is an important paradigm of statistical inference. The most popular and frequently used tests are the likelihood ratio, score and Wald's tests. However these classical test procedures suffer from the fact they are highly non-robust and hence may lead to fallacious conclusions in the presence of outliers. 
In this section, we define a class of robust tests for hypothesis regarding the ATE, using the proposed Mean-MDPDE of the ATE, 
so that these tests remain stable under arbitrary small departures from the specified (parametric) model distributions.
Proofs of all the results of this section are given in Appendix \ref{APP:Proof_Tests}.

\subsection{Testing for a specific value of the ATE}

Consider first the simplest problem of testing the simple null hypothesis $H_0: \Delta_1=\Delta_{10}$, where $\Delta_{10} \in \Theta \subseteq \mathbb{R}$. 
Although tests based on MLEs have several nice optimum properties, they are highly non-robust in presence of outliers in the sample. 
To circumvent this issue of non-robustness, we base our test statistic on the proposed Mean-MDPDE $\widehat{\Delta}_1^{(\alpha)}$ of the ATE, 
with tuning parameter $\alpha>0$. Noting that  $\sqrt{\widehat{\Sigma}(\alpha)/T_2} $ is the standard error of the Mean-MDPDE, 
we define the one-sample  test statistic for testing $H_0:\Delta_1=\Delta_{10}$ as given by
\begin{equation}\label{4.1}
W_1^{(\alpha)}=\frac{\sqrt{T_2}(\widehat{\Delta}_1^{(\alpha)} -\Delta_{10})}{{\sqrt{\widehat{\Sigma}(\alpha)}}}
\end{equation}
%In order to perform any statistical test, we first need to derive the asymptotic distribution of the test statistic under $H_0$. 
Using the asymptotic properties of the Mean-MDPDE presented in Section \ref{SEC:Asymp}, 
we can easily see that the asymptotic null distribution of the proposed test statistic $W_1^{(\alpha)}$ is indeed $\mathcal{N}(0,1)$.
So, denoting $z_\tau$ to be the upper $100(1-\tau)$-th quantile of the standard normal distribution, 
our rejection regions at the level of significance $\tau$, against different alternative hypotheses, would be as follow.   
\begin{enumerate}
	\item We reject $H_0$ against $H_1: \Delta_1>\Delta_{10}$ if $W_1^{(\alpha)}>z_\tau$
	\item We reject $H_0$ against $H_1: \Delta_1<\Delta_{10}$ if $W_1^{(\alpha)}<z_{1-\tau}$
	\item Reject $H_0$ against $H_1: \Delta_1\neq \Delta_{10}$ if $|W_1^{(\alpha)}|>z_{\tau/2}$
\end{enumerate}

\noindent
Next we shall derive an approximate power function $\beta_{W_1}^{(\alpha)}$ of the proposed test statistic 
in the following theorem which would, in turn, prove the consistency of the proposed test. 
In the following, we present results only for the first alternative  $H_1: \Delta_1>\Delta_{10}$; similar results can also be derived for the other two cases.

\begin{theorem}\label{THM:Power_Approx1}
Under the assumptions of Theorem \ref{THM:Asymp_ATE}, 
an approximation to the power function $\beta_{W_1^{(\alpha)}} $ for testing $H_0:\Delta_1=\Delta_{10}$ against $H_1: \Delta_1>\Delta_{10}$ 
using the test statistic $W_1^{(\alpha)}$ is given by
\begin{equation}\label{4.3}
\beta_{W_1}^{(\alpha)}(\Delta_1^*) \cong 1-\Phi \left(z_{\tau}-\sqrt{ \frac{T_2}{\widehat{\Sigma}(\alpha)}}\left(\Delta^*_1-\Delta_{10}\right)\right),
~~~~~~~~\mbox{ for }~ \Delta^*_1 > \Delta_{10},
\end{equation}
where $\tau$ is the significance level and $\Phi{(\cdot)}$ is the standard normal distribution function.
\end{theorem}

\medskip
\begin{corollary}
Under the assumptions of Theorem \ref{THM:Power_Approx1}, it follows from Equation \eqref{4.3} that 
$\lim\limits_{T_2\to \infty}\beta_{W_1}^{(\alpha)}(\Delta_1)= 1$, for all $\alpha >0$. 
Therefore, the proposed test is consistent in the sense of Fraser \hyperlink{ref8}{[8]}.
\end{corollary}

In order to produce a nontrivial asymptotic power, we can consider contiguous alternative hypotheses given by
\begin{equation} \label{4.4}
	H_{1,T_2}: \Delta_{1,T_2} = \Delta_{10}+T_2^{-1/2}d, 
\end{equation}
where $d \in \mathbb{R}\setminus\{0\}$ such that $\Delta_{1,T_2} \in \Theta$. 
The following theorem presents the asymptotic distribution of the test statistics under $H_{1,T_2}$ 
and subsequently the expression of asymptotic contiguous power.

\begin{theorem}\label{THM:Power_Contig1}
Under the assumptions of Theorem \ref{THM:Asymp_ATE}, we have the following results for testing $H_0:\Delta_1=\Delta_{10}$ against $H_1: \Delta_1>\Delta_{10}$ 
using the test statistic $W_1^{(\alpha)}$ at the level of significance $\tau$. 
\begin{itemize}
	\item The asymptotic distribution of $W_1^{(\alpha)}$ under the contiguous alternative $H_{1,T_2}$ in \eqref{4.4} is 
	normal with mean $d/\sqrt{\Sigma(\alpha)}$ and variance 1.
	
	\item The asymptotic power function $\overline{\beta}_{W_1}^{(\alpha)}$ at the contiguous sequence of alternatives in \eqref{4.4} is given by
	\begin{equation}
		\overline{\beta}_{W_1}^{(\alpha)}\left(d\right) = 1-\Phi \left(z_\tau-\frac{d}{\sqrt{\Sigma(\alpha)} }\right).
	\end{equation}
\end{itemize}
\end{theorem}

\subsection{Comparing the ATEs from Two Independent Studies}

The problem of comparing ATEs arises when we want to study the impact of a social policy on two independent populations. 
Examples may include the impact of government education policy among school-going children in urban and rural sectors. 
The primary interest in such cases is to test whether the difference between the two ATEs is statistically significant or not, i.e.,
we want to test the two-sample hypothesis $H_0:\Delta_{11}=\Delta_{12}$, where $\Delta_{1j}\in \mathbb{R}$ denote the ATE in the $j$-th population for $j=1,2$.  
We use their corresponding Mean-MDPDEs to construct a robust test for $H_0$. 

Let $T_{1j}$ and $T_{2j}$ be the number of observations available in pre- and post-treatment periods, respectively, for the $j$-th population
and $\widehat{\Delta}_{1j}^{(\alpha)}$ be the Mean-MDPDE of the ATE $\Delta_{1j}$ with tuning parameter $\alpha>0$ for each $j=1, 2$. 
Assuming identifiability of the model family, the difference between the two estimators $\widehat{\Delta}_{11}^{(\alpha)}$ and $\widehat{\Delta}_{12}^{(\alpha)}$ 
gives us an idea of the difference between the ATEs in the two samples and, hence, indicate any departure from the null hypothesis.
Denoting the standard error of $\widehat{\Delta}_{1j}^{(\alpha)}$ by $\sqrt{\widehat{\boldsymbol{\Sigma}_j}(\alpha)/T_{2j}}$ for $j=1, 2$, 
the two-sample  robust test statistic for testing $H_0:\Delta_{11}=\Delta_{12}$ can be defined as
\begin{equation}\label{4.7}
	W_2^{(\alpha)}=\frac{\left(\widehat{\Delta}_{11}^{(\alpha)}-\widehat{\Delta}_{12}^{(\alpha)}\right)}{
						\sqrt{{\frac{\widehat{\boldsymbol{\Sigma}_1}(\alpha)}{T_{21}}}+\frac{\widehat{{\Sigma}_2}(\alpha)}{T_{22}}}}.
\end{equation}

If the assumptions of Theorem \ref{THM:Asymp_ATE} hold for both the populations, then the two Mean-MDPDE estimators are asymptotically normally distributed
and, as a consequence, one can easily show that the asymptotic distribution of $W_2^{(\alpha)}$ under $H_0:\Delta_{11}=\Delta_{12}$ is indeed standard normal.
So, we reject $H_0$ against $H_1: \Delta_{11}>\Delta_{12}$, at a level of significance $\tau$, if $W_2^{(\alpha)}>z_\tau$. 
Tests for other alternatives can also be constructed in a routine manner.

As in the case of one sample problem, we can also derive an approximate power function $\beta_{W_2}^{(\alpha)}$ 
of the proposed two-sample test based on $W_2^{(\alpha)}$, which is presented in the following theorem.

\begin{theorem}\label{THM:Power_Approx2}
If the assumptions of Theorem \ref{THM:Asymp_ATE} hold for both populations, 
an approximation to the power function $\beta_{W_2}^{(\alpha)}$ for testing $H_0:\Delta_{11}=\Delta_{12}$ against $H_1: \Delta_{11}>\Delta_{12}$
using the test statistic $W_2^{(\alpha)}$ is given by
\begin{equation}\label{4.9}
	\beta_{W_2}^{(\alpha)}\left(\Delta^{*}_{11}, \Delta^{*}_{12}\right) \cong 1 - \Phi 
	\left(z_\tau-\frac{(\Delta^*_{11}-\Delta^*_{12})}{\sqrt{\frac{\widehat{\boldsymbol{\Sigma}_1}(\alpha)}{T_{21}}+\frac{\widehat{{\Sigma}_2}(\alpha)}{T_{22}}}}\right)
		~~~~~~~~\mbox{ for }~ \Delta^{*}_{11}>\Delta^{*}_{12}.
\end{equation}
%where $\tau$ is the significance level and $\Phi{()}$ is the standard normal distribution function.
\end{theorem} 

We can again see from Theorem \ref{THM:Power_Approx2} that
$$
\lim_{T_{21},T_{22}\to +\infty}\beta_{W_2}^{(\alpha)}(\Delta^*_{11},\Delta^*_{12})= 1, ~~~~~\mbox{ for any } ~\Delta^*_{11} > \Delta^*_{12}, ~~\alpha>0. 
$$
Therefore, the proposed two-sample test based on $W_2^{(\alpha)}$ is also asymptotically consistent.
We can, however, study its nontrivial asymptotic power by considering contiguous alternative hypotheses of a general form given by
\begin{equation} \label{4.10}
H_{1,d}:\Delta_{11}^d-\Delta_{12}^d = \sqrt{\frac{T_{21}+T_{22}}{T_{21}T_{22}}}d, ~~~~~~~~~~~ d \in \mathbb{R}\setminus\{0\}.
\end{equation}

\begin{theorem}\label{THM:Power_Contig2}
If the assumptions of Theorem \ref{THM:Asymp_ATE} hold for both populations, 
we have the following results for testing $H_0:\Delta_{11}=\Delta_{12}$ against $H_1: \Delta_{11}>\Delta_{12}$
	using the test statistic $W_2^{(\alpha)}$ at the level of significance $\tau$. 
Here, we assume $\eta=\displaystyle\lim_{T_{21},T_{22}\to +\infty} \frac{T_{22}}{T_{21}+T_{22}}$ exists having vale in $(0,1)$.
	\begin{itemize}
		\item The asymptotic distribution of $W_2^{(\alpha)}$ under the contiguous alternative $H_{1,d}$ in \eqref{4.10} is 
		normal with mean $d/\sqrt{\eta\Sigma_1(\alpha)+(1-\eta)\Sigma_2(\alpha)}$  and variance 1, as $T_{21},T_{22}\to +\infty$.
		
		\item The asymptotic power function $\overline{\beta}_{W_2}^{(\alpha)}$ at the contiguous sequence of alternatives in \eqref{4.10}, 
		as $T_{21},T_{22}\to +\infty$, is given by
		\begin{equation}
			\overline{\beta}_{W_2}^{(\alpha)}\left(d\right) = 1-\Phi\left(z_\tau-\frac{d}{\sqrt{\eta\Sigma_1(\alpha)+(1-\eta)\Sigma_2(\alpha)}}\right).
		\end{equation}
	\end{itemize}
\end{theorem}

\section{Simulation Studies}
\label{SEC:SIMULATION}

\subsection{Performance of the Mean-MDPDE of ATE}

We have conducted extensive simulation studies to examine the finite-sample performances of our proposed Mean-MDPDE of the ATE
by examining their empirical bias and mean-square error (MSE) under both uncontaminated (pure) as well as contaminated data.
Additionally, for comparison purpose, we also compute the same error measures for existing common ATE estimators, namely the HCW estimator \cite{Hsiao/etc:2012}, 
the DID estimator \cite[][Eq.~(2.6)]{Li/Bell:2017a}, the augmented DID (ADID) estimator \cite[][Eq.~(2.15)]{Li/Bell:2017a} and 
the MSCM estimator \cite[][Eq.~(2.4)]{Li:2020} which are then compared with our proposal in terms of robustness and efficiency. 
\\

%To assess for performance of the methods, we compute the MSE which is defined as $$ MSE = \frac{1}{M}\sum_{j=1}^{M}[\widehat{\Delta}_{1,j}^{(\alpha)}-\bar{\Delta}_1]^2$$ and the bias which is defined as $$Bias = \frac{1}{M}\sum_{j=1}^{M}[\widehat{\Delta}_{1,j}^{(\alpha)}-\bar{\Delta}_1]$$ where $M$ = 2,500 is the number of replications and $j$ denotes the estimation result for the $j$th replication.
%Also, $\widehat{\Delta}_{1,j}^{(\alpha)}=T_2^{-1}\sum_{t=T_1+1}^{T}\widehat{\Delta}_{1t.j}^{(\alpha)}$  for $j=1,...,M$

\noindent\textbf{Simulation Set-up:}
We consider the simulation experiments following \cite{Li/Bell:2017b}, 
where one treatment and two control units ($N=3$) are observed over $T$ time-points and the first unit got the treatment at time-point $T_1+1$. 
To generate the observations according to the HCW factor model, we first simulate one unobserved latent variable ($K=1$)
from an AR(1) process as $f_t$ = $0.5f_{t-1} + v_t$ with $v_t$ being independently distributed as $N(0,1)$ for each $t=1, \ldots, T$. 
Then the observations of the three units are generated via $\boldsymbol{y}_t= (y_{1t}, y_{2t},y_{3t})' = \textbf{a} + \textbf{b} f_t + \textbf{u}_t$, 
for every $t= 1,...,T$, where $\textbf{a}= (1,1,1)'$, $\textbf{b}=(1,1,1)'$ and $\textbf{u}_t= (u_{1t}, u_{2t},u_{3t})'$ 
with $u_{jt}\sim \mathcal{N}(0,1)$ independently for each $j=1,2,3$. 
To incorporate the treatment effects on the first unit, we modify it as  $y_{1t} \leftarrow y_{1t} + \Delta_{1t}$ 
for the post-treatment time points $t=T_1+1, \ldots, T_2$, 
where $\Delta_{1t}$s are the individual treatment effects at time $t$ and are generated by 
$\Delta_{1t}= \exp(z_t)/[1+\exp(z_t)]+1$ with $z_t=0.5z_{t-1}+\epsilon_t$ and $\epsilon_t\sim \mathcal{N}(0,0.25)$. 
These errors ($\epsilon_t$s) are generated independent of each other and also independent of $(\widetilde{y}_t, u_{jt})$ for all $t= 1,...T$ and $j=1,2,3$. 
Then, we estimate the ATE from the sample data (using different estimators) and compute the empirical bias and MSE based on 2500 replications
and against the true ATE value  $\bar{\Delta}_{1}=T_2^{-1}\sum_{t=T_1+1}^{T}\Delta_{1t}$. 

Additionally, in order to study the robustness of the estimators, we repeat the above simulation exercises 
by contaminating 5\%  or 20\% of the sample observations on the first unit (treated response).
These contaminated observations (randomly selected from the pool) are generated by using $u_{1t} \sim \mathcal{N}(5,1)$  instead of $\mathcal{N}(0,1)$
either in the pre-treatment or in the post-treatment periods. \\

%The different choices of $u_{jt}$ considered are as follows:- \\
%\begin{enumerate}
%	\item [1]$u_{jt}$ as iid $N(0,1)$.\\
%	\item[2] $u_{jt}$ as iid $N(5,1)$ for $5\%$ of the pre treatment $y_{1t}$. For the rest, $u_{jt}$ as iid $N(0,1)$ was taken.\\
%	\item[3]$u_{jt}$ as iid $N(5,1)$ for $20\%$ of the pre treatment $y_{1t}$. For the rest, $u_{jt}$ as iid $N(0,1)$ was taken.\\
%	%4) $u_{jt}$ as iid $N(5,1)$ for $5\%$ of the pre treatment $y_{2t}$ For the rest, $u_{jt}$ as iid $N(0,1)$ was taken.\\
%	%5) $u_{jt}$ as iid $N(5,1)$ for $20\%$ of the pre treatment $y_{2t}$ For the rest, $u_{jt}$ as iid $N(0,1)$ was taken.\\
%	\item[4]$u_{jt}$ as iid $N(5,1)$ for $5\%$ of the post treatment $y_{1t}$. For the rest, $u_{jt}$ as iid $N(0,1)$ was taken.\\
%	\item[5]$u_{jt}$ as iid $N(5,1)$ for $20\%$ of the post treatment $y_{1t}$. For the rest, $u_{jt}$ as iid $N(0,1)$ was taken.\\
%	%8)$u_{jt}$ as iid $N(5,1)$ for $5\%$ of the post treatment $y_{2t}$ For the rest, $u_{jt}$ as iid $N(0,1)$ was taken.\\
%	%9) $u_{jt}$ as iid $N(5,1)$ for $20\%$ of the post treatment $y_{2t}$ For the rest, $u_{jt}$ as iid $N(0,1)$ was taken.\\
%\end{enumerate}

\begin{table}
	\caption{Empirical biases of the ATE estimators under pure and contaminated data} 
	\centering
	\color{black}
	\begin{adjustbox}{height=0.49\textheight}
		\begin{tabular}{|l|cc|cc|cc|cc|cc|}
			\hline
	$T_1$	&	\multicolumn{2}{|c|}{20}			&	\multicolumn{2}{|c|}{50}			&	\multicolumn{2}{|c|}{100}			&	\multicolumn{2}{|c|}{200}			&	\multicolumn{2}{|c|}{400}			\\
$T_2$	&	20	&	80	&	20	&	80	&	20	&	80	&	20	&	80	&	20	&	80	\\
\hline																					
\multicolumn{11}{|c|}{\textbf{Pure data}}\\																					
HCW 	&	-0.008	&	-0.006	&	-0.002	&	-0.001	&	0.012	&	0.002	&	0.016	&	-0.002	&	0.007	&	0.002	\\
DID	&	0.006	&	0.007	&	0.000	&	0.004	&	0.000	&	0.003	&	-0.011	&	-0.003	&	-0.005	&	0.002	\\
ADID	&	0.001	&	-0.003	&	0.015	&	-0.006	&	-0.008	&	-0.006	&	0.007	&	-0.007	&	0.000	&	0.004	\\
MSCM 	&	0.093	&	0.114	&	0.040	&	0.035	&	0.013	&	0.014	&	0.016	&	0.002	&	0.005	&	0.009	\\
Mean-MDPDE \\																					
~~	$\alpha=0.1$	&	-0.007	&	-0.005	&	-0.003	&	-0.001	&	0.011	&	0.002	&	0.015	&	-0.002	&	0.007	&	0.002	\\
~~	$\alpha=0.3$	&	-0.003	&	-0.002	&	-0.004	&	-0.002	&	0.010	&	0.002	&	0.014	&	-0.002	&	0.007	&	0.002	\\
~~	$\alpha=0.5$	&	0.000	&	0.001	&	-0.005	&	-0.003	&	0.010	&	0.001	&	0.014	&	-0.002	&	0.006	&	0.003	\\
~~	$\alpha=0.7$	&	-0.001	&	0.003	&	-0.006	&	-0.003	&	0.009	&	0.001	&	0.013	&	-0.002	&	0.006	&	0.003	\\
~~	$\alpha=1$	&	0.002	&	0.009	&	-0.008	&	-0.004	&	0.008	&	0.000	&	0.012	&	-0.002	&	0.006	&	0.003	\\
\hline																					
\multicolumn{11}{|c|}{\textbf{$5\%$ Contamination in pre treatment period}}\\																					
HCW   	&	-0.252	&	-0.250	&	-0.241	&	-0.245	&	-0.245	&	-0.242	&	-0.240	&	-0.252	&	-0.247	&	-0.244	\\
DID	&	-0.256	&	-0.245	&	-0.252	&	-0.243	&	-0.239	&	-0.248	&	-0.244	&	-0.243	&	-0.249	&	-0.249	\\
ADID	&	-0.248	&	-0.251	&	-0.242	&	-0.246	&	-0.244	&	-0.242	&	-0.240	&	-0.252	&	-0.247	&	-0.244	\\
MSCM	&	-0.155	&	-0.155	&	-0.206	&	-0.200	&	-0.225	&	-0.232	&	-0.242	&	-0.243	&	-0.247	&	-0.250	\\
Mean-MDPDE \\																					
~~	$\alpha=0.1$   	&	-0.157	&	-0.183	&	-0.163	&	-0.172	&	-0.154	&	-0.151	&	-0.145	&	-0.157	&	-0.152	&	-0.149	\\
~~	$\alpha=0.3$	&	-0.047	&	-0.063	&	-0.046	&	-0.055	&	-0.042	&	-0.040	&	-0.033	&	-0.046	&	-0.041	&	-0.038	\\
~~	$\alpha=0.5$	&	-0.009	&	-0.022	&	-0.015	&	-0.024	&	-0.015	&	-0.013	&	-0.005	&	-0.018	&	-0.014	&	-0.012	\\
~~	$\alpha=0.7$	&	0.002	&	-0.009	&	-0.006	&	-0.015	&	-0.007	&	-0.005	&	0.003	&	-0.010	&	-0.006	&	-0.004	\\
~~	$\alpha=1$	&	0.007	&	-0.007	&	-0.001	&	-0.012	&	-0.003	&	-0.002	&	0.008	&	-0.006	&	-0.002	&	0.000	\\
\hline																					
\multicolumn{11}{|c|}{\textbf{$20\%$ Contamination in pre treatment period}}\\																				
HCW     	&	-1.013	&	-1.003	&	-1.000	&	-0.992	&	-0.991	&	-0.997	&	-0.992	&	-1.001	&	-0.997	&	-0.992	\\
DID	&	-1.001	&	-0.994	&	-1.006	&	-1.002	&	-0.987	&	-1.000	&	-0.994	&	-0.992	&	-0.999	&	-1.002	\\
ADID	&	-1.012	&	-1.002	&	-1.000	&	-0.991	&	-0.991	&	-0.997	&	-0.993	&	-1.000	&	-0.996	&	-0.992	\\
MSCM	&	1-0.8913	&	-0.868	&	-0.957	&	-0.958	&	-0.975	&	-0.981	&	-0.993	&	-0.998	&	-1.005	&	-1.004	\\
Mean-MDPDE \\																					
~~	$\alpha=0.1$   	&	-0.885	&	-0.911	&	-0.895	&	-0.901	&	-0.882	&	-0.887	&	-0.881	&	-0.889	&	-0.884	&	-0.879	\\
~~	$\alpha=0.3$	&	-0.650	&	-0.670	&	-0.645	&	-0.651	&	-0.627	&	-0.631	&	-0.625	&	-0.633	&	-0.627	&	-0.622	\\
~~	$\alpha=0.5$	&	-0.412	&	-0.435	&	-0.349	&	-0.358	&	-0.301	&	-0.303	&	-0.282	&	-0.294	&	-0.275	&	-0.269	\\
~~	$\alpha=0.7$	&	-0.257	&	-0.282	&	-0.164	&	-0.179	&	-0.115	&	-0.113	&	-0.087	&	-0.103	&	-0.083	&	-0.081	\\
~~	$\alpha=1$	&	-0.143	&	-0.167	&	-0.065	&	-0.076	&	-0.041	&	-0.038	&	-0.026	&	-0.042	&	-0.032	&	-0.031	\\
\hline																					
\multicolumn{11}{|c|}{\textbf{$5\%$ Contamination in post treatment period}}\\																				
HCW     	&	0.251	&	0.243	&	0.245	&	0.254	&	0.248	&	0.249	&	0.256	&	0.252	&	0.255	&	0.253	\\
DID	&	0.236	&	0.255	&	0.268	&	0.247	&	0.242	&	0.253	&	0.258	&	0.247	&	0.253	&	0.252	\\
ADID	&	0.251	&	0.243	&	0.246	&	0.253	&	0.248	&	0.249	&	0.256	&	0.251	&	0.255	&	0.253	\\
MSCM	&	0.372	&	0.352	&	0.298	&	0.286	&	0.266	&	0.267	&	0.272	&	0.265	&	0.242	&	0.249	\\
Mean-MDPDE \\																					
~~	$\alpha=0.1$   	&	0.251	&	0.243	&	0.245	&	0.254	&	0.248	&	0.249	&	0.256	&	0.251	&	0.255	&	0.253	\\
~~	$\alpha=0.3$	&	0.250	&	0.243	&	0.245	&	0.254	&	0.248	&	0.248	&	0.256	&	0.251	&	0.255	&	0.253	\\
~~	$\alpha=0.5$	&	0.249	&	0.243	&	0.245	&	0.254	&	0.247	&	0.248	&	0.256	&	0.251	&	0.255	&	0.253	\\
~~	$\alpha=0.7$	&	0.249	&	0.244	&	0.246	&	0.254	&	0.247	&	0.248	&	0.256	&	0.250	&	0.255	&	0.253	\\
~~	$\alpha=1$	&	0.247	&	0.240	&	0.246	&	0.255	&	0.246	&	0.248	&	0.256	&	0.250	&	0.255	&	0.253	\\
\hline																					
\multicolumn{11}{|c|}{\textbf{$20\%$ Contamination in post treatment period}}\\																				
HCW     	&	1.015	&	0.993	&	0.993	&	1.005	&	0.997	&	0.998	&	1.020	&	1.000	&	1.010	&	1.009	\\
DID 	&	0.980	&	1.007	&	1.014	&	1.000	&	1.004	&	1.001	&	1.005	&	0.998	&	1.012	&	0.995	\\
ADID 	&	1.015	&	0.993	&	0.994	&	1.004	&	0.997	&	0.998	&	1.020	&	1.000	&	1.011	&	1.009	\\
MSCM 	&	1.116	&	1.104	&	1.044	&	1.035	&	1.016	&	1.016	&	1.012	&	1.015	&	0.984	&	1.001	\\
Mean-MDPDE \\																					
~~	$\alpha=0.1$   	&	1.015	&	0.993	&	0.993	&	1.005	&	0.997	&	0.998	&	1.020	&	1.000	&	1.010	&	1.009	\\
~~	$\alpha=0.3$	&	1.014	&	0.993	&	0.993	&	1.005	&	0.996	&	0.997	&	1.020	&	1.000	&	1.011	&	1.009	\\
~~	$\alpha=0.5$	&	1.013	&	0.993	&	0.993	&	1.005	&	0.996	&	0.997	&	1.020	&	0.999	&	1.010	&	1.009	\\
~~	$\alpha=0.7$	&	1.013	&	0.993	&	0.993	&	1.005	&	0.995	&	0.997	&	1.020	&	0.999	&	1.010	&	1.009	\\
~~	$\alpha=1$	&	1.011	&	0.990	&	0.993	&	1.006	&	0.995	&	0.996	&	1.020	&	0.999	&	1.010	&	1.009	\\
\hline																							
\end{tabular}
	\end{adjustbox}
\label{TAB:Bias}
\end{table}

\begin{table}
	\caption{Empirical MSEs of the ATE estimators under pure and contaminated data} 
	\centering
	\color{black}
	\begin{adjustbox}{height=0.49\textheight}
		\begin{tabular}{|l|cc|cc|cc|cc|cc|}
			\hline
	$T_1$	&	\multicolumn{2}{|c|}{20}			&	\multicolumn{2}{|c|}{50}			&	\multicolumn{2}{|c|}{100}			&	\multicolumn{2}{|c|}{200}			&	\multicolumn{2}{|c|}{400}			\\
$T_2$	&	20	&	80	&	20	&	80	&	20	&	80	&	20	&	80	&	20	&	80	\\
\hline																					
\multicolumn{11}{|c|}{\textbf{Pure data}}\\																					
HCW 	&	0.187	&	0.114	&	0.115	&	0.053	&	0.095	&	0.039	&	0.082	&	0.026	&	0.081	&	0.023	\\
DID	&	0.145	&	0.095	&	0.104	&	0.045	&	0.089	&	0.034	&	0.086	&	0.024	&	0.075	&	0.022	\\
ADID	&	0.172	&	0.106	&	0.116	&	0.053	&	0.097	&	0.035	&	0.086	&	0.028	&	0.078	&	0.024	\\
MSCM 	&	0.220	&	0.161	&	0.127	&	0.064	&	0.108	&	0.039	&	0.093	&	0.030	&	0.083	&	0.025	\\
Mean-MDPDE \\																					
~~	$\alpha=0.1$	&	0.188	&	0.115	&	0.116	&	0.053	&	0.095	&	0.039	&	0.082	&	0.026	&	0.081	&	0.023	\\
~~	$\alpha=0.3$	&	0.197	&	0.125	&	0.119	&	0.056	&	0.097	&	0.040	&	0.083	&	0.027	&	0.082	&	0.023	\\
~~	$\alpha=0.5$	&	0.220	&	0.142	&	0.124	&	0.061	&	0.099	&	0.042	&	0.084	&	0.028	&	0.082	&	0.023	\\
~~	$\alpha=0.7$	&	0.246	&	0.166	&	0.130	&	0.067	&	0.101	&	0.044	&	0.085	&	0.029	&	0.083	&	0.024	\\
~~	$\alpha=1$	&	0.286	&	0.206	&	0.141	&	0.078	&	0.105	&	0.048	&	0.087	&	0.030	&	0.084	&	0.024	\\
\hline																					
\multicolumn{11}{|c|}{\textbf{$5\%$ Contamination in pre treatment period}}\\																					
HCW   	&	0.344	&	0.250	&	0.197	&	0.143	&	0.174	&	0.107	&	0.155	&	0.097	&	0.147	&	0.086	\\
DID	&	0.286	&	0.215	&	0.186	&	0.134	&	0.158	&	0.106	&	0.144	&	0.090	&	0.145	&	0.087	\\
ADID	&	0.315	&	0.239	&	0.195	&	0.141	&	0.171	&	0.107	&	0.154	&	0.097	&	0.146	&	0.086	\\
MSCM	&	0.280	&	0.215	&	0.187	&	0.124	&	0.156	&	0.101	&	0.152	&	0.091	&	0.141	&	0.088	\\
Mean-MDPDE \\																					
~~	$\alpha=0.1$   	&	0.268	&	0.204	&	0.165	&	0.101	&	0.133	&	0.068	&	0.116	&	0.057	&	0.107	&	0.047	\\
~~	$\alpha=0.3$	&	0.230	&	0.158	&	0.132	&	0.064	&	0.105	&	0.042	&	0.093	&	0.032	&	0.085	&	0.026	\\
~~	$\alpha=0.5$	&	0.236	&	0.158	&	0.132	&	0.062	&	0.104	&	0.041	&	0.092	&	0.031	&	0.083	&	0.024	\\
~~	$\alpha=0.7$	&	0.257	&	0.175	&	0.137	&	0.067	&	0.105	&	0.043	&	0.093	&	0.031	&	0.084	&	0.025	\\
~~	$\alpha=1$	&	0.302	&	0.203	&	0.147	&	0.075	&	0.109	&	0.047	&	0.094	&	0.033	&	0.084	&	0.025	\\
\hline																					
\multicolumn{11}{|c|}{\textbf{$20\%$ Contamination in pre treatment period}}\\																					
HCW     	&	1.513	&	1.370	&	1.219	&	1.132	&	1.126	&	1.071	&	1.096	&	1.049	&	1.088	&	1.018	\\
DID	&	1.355	&	1.286	&	1.193	&	1.134	&	1.106	&	1.074	&	1.084	&	1.030	&	1.088	&	1.036	\\
ADID	&	1.480	&	1.342	&	1.210	&	1.127	&	1.122	&	1.070	&	1.096	&	1.048	&	1.086	&	1.017	\\
MSCM	&	1.249	&	1.105	&	1.159	&	1.075	&	1.130	&	1.054	&	1.134	&	1.054	&	1.144	&	1.054	\\
Mean-MDPDE \\																					
~~	$\alpha=0.1$   	&	1.266	&	1.223	&	1.027	&	0.964	&	0.926	&	0.870	&	0.890	&	0.841	&	0.877	&	0.807	\\
~~	$\alpha=0.3$	&	0.998	&	0.942	&	0.678	&	0.616	&	0.561	&	0.500	&	0.515	&	0.462	&	0.494	&	0.426	\\
~~	$\alpha=0.5$	&	0.791	&	0.725	&	0.404	&	0.339	&	0.273	&	0.210	&	0.212	&	0.156	&	0.183	&	0.118	\\
~~	$\alpha=0.7$	&	0.658	&	0.608	&	0.256	&	0.195	&	0.152	&	0.089	&	0.110	&	0.053	&	0.095	&	0.035	\\
~~	$\alpha=1$	&	0.565	&	0.489	&	0.190	&	0.120	&	0.119	&	0.056	&	0.096	&	0.037	&	0.087	&	0.027	\\
\hline																					
\multicolumn{11}{|c|}{\textbf{$5\%$ Contamination in post treatment period}}\\																					
HCW     	&	0.310	&	0.189	&	0.232	&	0.134	&	0.210	&	0.115	&	0.211	&	0.105	&	0.217	&	0.102	\\
DID	&	0.273	&	0.175	&	0.228	&	0.121	&	0.206	&	0.111	&	0.210	&	0.102	&	0.203	&	0.102	\\
ADID	&	0.300	&	0.185	&	0.230	&	0.133	&	0.210	&	0.115	&	0.211	&	0.104	&	0.218	&	0.102	\\
MSCM	&	0.419	&	0.288	&	0.278	&	0.157	&	0.234	&	0.128	&	0.230	&	0.116	&	0.207	&	0.101	\\
Mean-MDPDE \\																					
~~	$\alpha=0.1$   	&	0.311	&	0.190	&	0.232	&	0.135	&	0.210	&	0.116	&	0.211	&	0.105	&	0.218	&	0.102	\\
~~	$\alpha=0.3$	&	0.319	&	0.200	&	0.233	&	0.137	&	0.211	&	0.117	&	0.211	&	0.105	&	0.218	&	0.102	\\
~~	$\alpha=0.5$	&	0.337	&	0.220	&	0.237	&	0.140	&	0.213	&	0.118	&	0.212	&	0.106	&	0.219	&	0.103	\\
~~	$\alpha=0.7$	&	0.363	&	0.243	&	0.243	&	0.145	&	0.215	&	0.120	&	0.213	&	0.107	&	0.219	&	0.103	\\
~~	$\alpha=1$	&	0.400	&	0.270	&	0.254	&	0.154	&	0.219	&	0.124	&	0.214	&	0.108	&	0.220	&	0.104	\\
\hline																					
\multicolumn{11}{|c|}{\textbf{$20\%$ Contamination in post treatment period}}\\																					
HCW     	&	1.413	&	1.156	&	1.298	&	1.115	&	1.282	&	1.085	&	1.336	&	1.077	&	1.315	&	1.088	\\
DID 	&	1.312	&	1.159	&	1.329	&	1.095	&	1.289	&	1.081	&	1.293	&	1.073	&	1.302	&	1.063	\\
ADID 	&	1.406	&	1.152	&	1.297	&	1.114	&	1.282	&	1.084	&	1.335	&	1.076	&	1.316	&	1.088	\\
MSCM 	&	1.661	&	1.415	&	1.435	&	1.180	&	1.332	&	1.115	&	1.315	&	1.112	&	1.248	&	1.079	\\
Mean-MDPDE \\																					
~~	$\alpha=0.1$   	&	1.414	&	1.157	&	1.298	&	1.116	&	1.282	&	1.085	&	1.336	&	1.076	&	1.315	&	1.088	\\
~~	$\alpha=0.3$	&	1.420	&	1.166	&	1.299	&	1.119	&	1.282	&	1.085	&	1.336	&	1.076	&	1.315	&	1.089	\\
~~	$\alpha=0.5$	&	1.435	&	1.185	&	1.303	&	1.123	&	1.283	&	1.086	&	1.337	&	1.076	&	1.316	&	1.089	\\
~~	$\alpha=0.7$	&	1.458	&	1.209	&	1.308	&	1.127	&	1.284	&	1.088	&	1.338	&	1.076	&	1.316	&	1.090	\\
~~	$\alpha=1$	&	1.494	&	1.231	&	1.321	&	1.136	&	1.287	&	1.091	&	1.340	&	1.077	&	1.317	&	1.091	\\
\hline																					
		\end{tabular}
	\end{adjustbox}
\label{TAB:MSE}
\end{table}

\noindent\textbf{Simulation Results:}
The resulting bias and MSE values of different ATE estimators are presented in Tables \ref{TAB:Bias} and \ref{TAB:MSE}, respectively, 
for $T_1= 100,200,400$ and $T_2 = T-T_1 = 20,80,320$. 
It is clearly seen that that, when there is no data contamination, the DID estimator has the least MSE followed by HCW estimator. 
But, our proposed Mean-MDPDE at smaller $\alpha>0$ are also quite competitive; 
their loss in efficiency (in terms of MSE) is not quite significant and further diminishes as the sample size increases. 
On the contrary, in the presence of contamination in pre-treatment periods, the proposed Mean-MDPD estimator(s) clearly outsmarts all the existing ATE estimators;
in this respect, Mean-MDPDE with larger $\alpha>0$ performs better indicating their increasing robustness properties with increase in $\alpha$. 

However, the mean-MDPDE fails to generate robust ATE estimators in the presence of contamination in post-treatment periods,
when it performs mostly at par with all existing estimators.
Such robustness behavior of our Mean-MDPDE of the ATE  will be further justified theoretically via the influence function analyses 
in Section \ref{SEC:IF}.

\begin{figure}
	\centering
	%------------------------------------------------------
	\subfloat[$T_1=100$,  $T_2=20$]{
		\includegraphics[width=0.5\textwidth, height=0.4\textwidth]{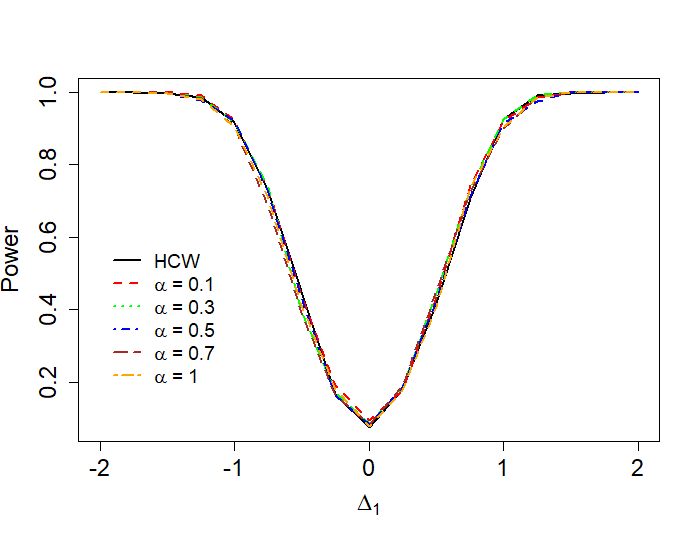}
		\includegraphics[width=0.5\textwidth, height=0.4\textwidth]{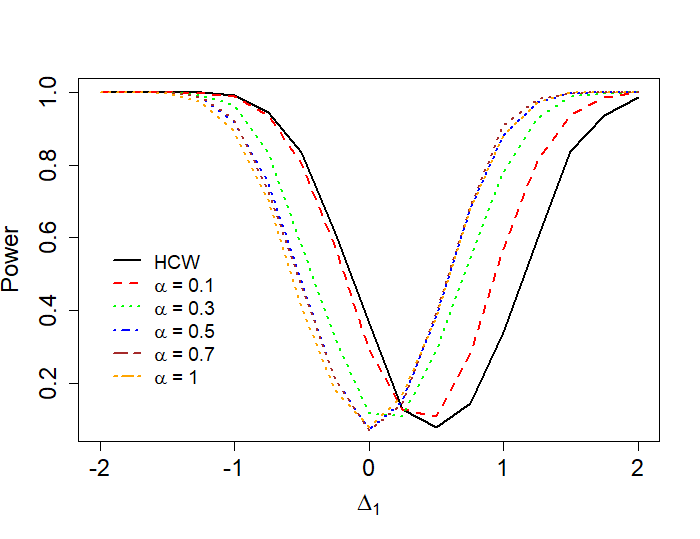}
		\label{FIG:IF_mu}}\\
	%--------------------------------------------------------------------
	\subfloat[$T_1=400$, $T_2=20$]{
		\includegraphics[width=0.5\textwidth, height=0.4\textwidth]{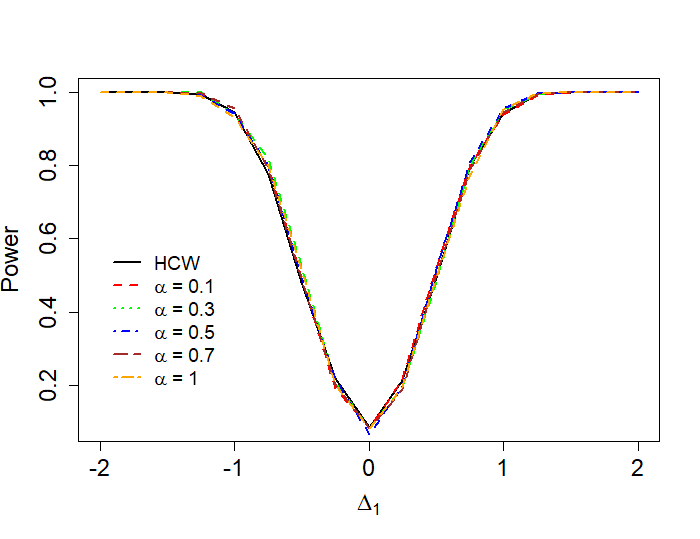}
		\includegraphics[width=0.5\textwidth, height=0.4\textwidth]{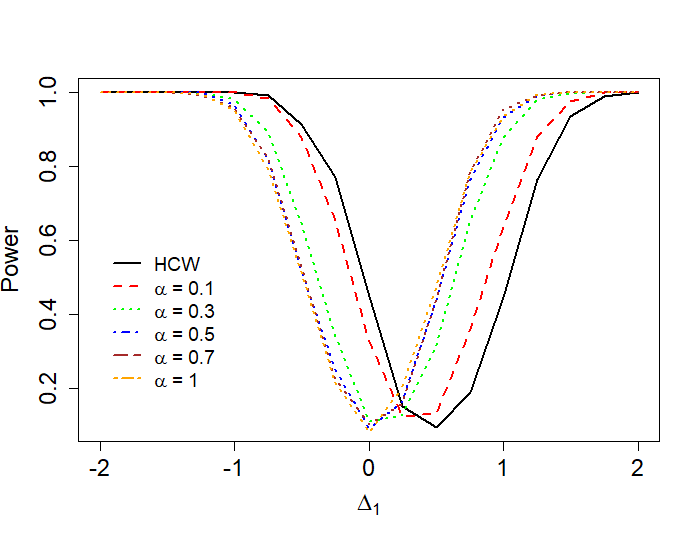}
		\label{FIG:IF_sigma}}\\
	%--------------------------------------------------------------------
	\subfloat[$T_1=400$, $T_2=100$]{
		\includegraphics[width=0.5\textwidth, height=0.4\textwidth]{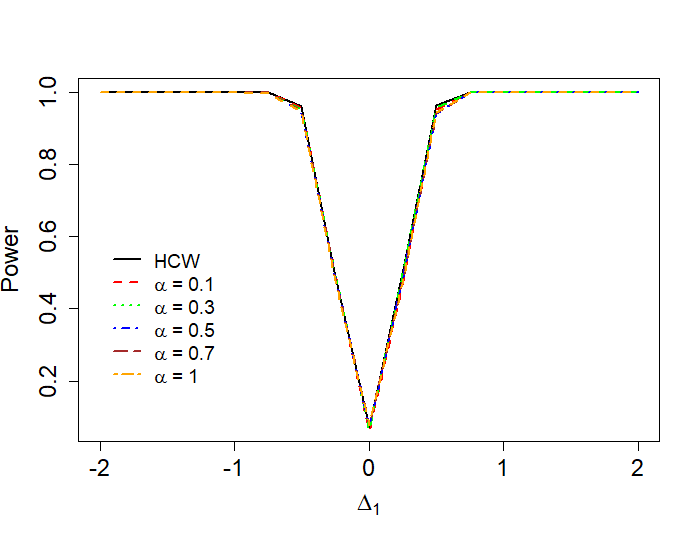}
		\includegraphics[width=0.5\textwidth, height=0.4\textwidth]{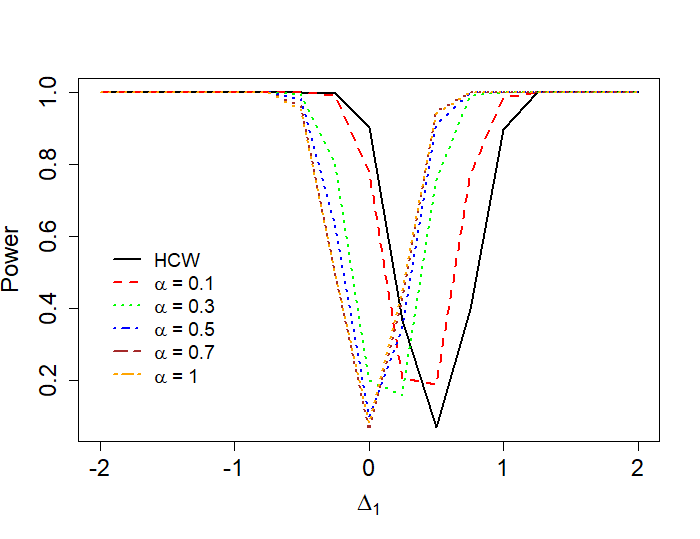}
		\label{FIG:IF_gamma}}
	%---------------------------------------------------------------------------
	\caption{Empirical power function for the proposed test based on the Mean-MDPDE of the ATE in the simulation
		with uncontaminated data (left panel) and with $10\%$ contamination in the pre treatment period (right panel).}
	\label{FIG:power}
\end{figure}

\subsection{Performance of the MDPDE based Tests}

We also repeat the same simulation study as described in the previous subsection to investigate the performances of the 
proposed robust Wald-type tests for testing significance of the ATE for different finite sample-sizes $(T_1, T_2)$. 
In particular, we test for the hypothesis  $H_0:$ $\Delta_1=0$ against $H_1:\Delta_1 \neq 0$
and plot the empirical power functions, obtained  based on $5000$ replications at the 5\% level of significance, for pure data 
as well as 10\% contaminated data in Figure \ref{FIG:power}. 
The contamination is taken in the pre-treatment period as in the previous subsection.
Note that the value of the power function at $\Delta_1=0$ yields the empirical size of the corresponding tests. 
For the sake of comparisons, the power function for the Wald-type test based on the HCW estimator is also plotted in the same Figure \ref{FIG:power}.

It can be seen from the figures that the robust MDPDE-based Wald type tests perform quite similarly to the classical test under pure data;
they all maintain the empirical size at the specified level and have powers increasing to one as the sample size increases 
or as the  alternative moves away from the null (i.e., $\Delta_1$ moves away from zero).
Under contamination, however, the classical HCW estimator becomes extremely non-robust
and hence the corresponding test have significantly inflated size and decreased power at any fixed alternative ($\Delta_1>0$) compared to the pure data case. 
On the contrary, the proposed MDPDE-based tests with higher values of $\alpha>0$ remain stable, both in terms of size and power,  
even under contamination  indicating their claimed robustness benefits.

{\cbl

\subsection{On Selection of the Tuning parameter $\alpha$}
\label{SEC:Sel_alpha}
}

It is important to note that the proposed ATE estimator and its performance crucially depend on a tuning parameter $\alpha$ $(0\leq \alpha \leq 1)$. 
By varying this tuning parameter, the strength of the estimator can be controlled via a trade-offs between efficiency and robustness of the resulting inference. 
Choice of $\alpha$ near 0 will produce efficiency close to that of the MLE but will lack in terms of robustness; 
on the other hand, the robustness increases with increase in $\alpha$.  
So, it is important to choose a proper $\alpha$ in any practical application of the proposed methodology. 
Based on our extensive simulations and empirical investigations, it has been observed that a choice of $\alpha\approx 0.3, 0.5$ provides the best trade-off
despite the amount of contamination in data (which is often unknown); 
so it can serve as our empirical recommendation for the choice of $\alpha$  in any practical application of the proposed methodology. 
However, a data-driven algorithm for choosing an optimum $\alpha$ values for a given dataset could be useful in some complex contexts.

{\cbl 
One such algorithm by minimizing  and estimate of the asymptotic mean squared error has been explored by \cite{Ghosh/Basu:2015}
in the context of linear regression models which can be directly adopted in our present context for estimating the ATE.
As an alternative way, suggested by a reviewer, here we have briefly explored the use of bootstrap 
for the selection of optimal tuning parameter value $\alpha$. 
Such a bootstrap based approach has recently been used for tuning parameter selection in many problems; see, e.g., \cite{Ozkale/Altuner:2021}.
However, the standard application of bootstrap requires the assumption of independence between the observations,
which is clearly not a valid assumption in our case with panel data. 
So, we resort to a more specialized approach to bootstrapping, called Moving Block Bootstrap \citep{Liu:1992}.
In this approach, for a sample of size $n$, we consider $n-b+1$ overlapping blocks of size $b$, 
each block containing $b$ consecutive data points. 
Then, instead of resampling from the observations as in usual bootstrap, we consider resampling the blocks instead,
to preserve the correlation structure of the original observations to some extent. 
Finally, these blocks are combined together to form a bootstrap sample of size $n$. 
Hence, our empirical data-driven algorithm for selection of $\alpha$ using this approach of moving block bootstrap 
can be summarized as follows:  
\begin{enumerate}
	\item Generate a resample of size $T_1$ from the pre-treatment period using moving block bootstrap. Repeat this process $B$ times to generate $B$ bootstrap samples.
	\item For the $j^{th}$ bootstrap sample, $j=1, \ldots, B$, compute the parameter estimates  for several choices of $\alpha$.
	\item Suppose $T^*$ unique observations are included in the bootstrap sample. Then, for the $j^{th}$ sample, use the obtained parameter estimates to get a robust estimate of ATE for the observations not included in the sample as: $$\hat{\Delta}_{1j, \text{boot}}^{(\alpha)}= \frac{1}{T_1-T^*} \sum_{t \in S} (y_{1t}-\hat{y}_{1jt,\text{boot}}^{(\alpha)})$$ where $S$ consists of $T_1-T^*$ time-points not included in the bootstrap sample. 
	\item Let $\alpha_{opt}$ denote the optimal choice of $\alpha$. Then, $\alpha_{opt}$ is given by: $$\alpha_{opt}= \underset{\alpha}{\text{arg min}}\left(\frac{1}{B} \sum_{j=1}^B \hat{\Delta}_{1j, \text{boot}}^{(\alpha)2}\right)$$
\end{enumerate}
The rationale behind such a selection criteria is that we do not expect to have any kind of treatment effect at any of the pre-treatment time points. Hence, we would like to make the mean squared bootstrap estimates of ATE as close to 0 as possible.

We have briefly illustrated  the applicability of this suggested procedure in case of our real data example in a later section.
However, considering the contents and length of the present manuscript, 
more detailed exploration and verification of the performances of this suggested ways of selection of $\alpha$ under different set-ups 
are kept for a future work. 
}

\section{Theoretical Study of Robustness: Influence Functions}
\label{SEC:IF}

Let us now further explore the robustness properties of the proposed MDPDE of the ATE theoretically via the classical influence function analysis. 
Let $G_{pre}(\boldsymbol{x},y_{1})$ and $G_{post}(\boldsymbol{x},y_{1})$ denote the joint distributions of $\boldsymbol{x}$ and $y_1$ 
in the pre-treatment and the post-treatment periods, respectively, where $\boldsymbol{x}=(1,\boldsymbol{\widetilde{y}})$. 
Then, the ATE estimator $\widehat{\Delta}_1^{(\alpha)}$ can be re-written in terms of their empirical estimator, namely  $G_{n,pre}$ and $G_{n,pre}$, respectively, 
as follows:
\begin{align}
	\widehat{\Delta}_1^{(\alpha)}=\frac{1}{T_2}\sum_{t=T_1+1}^{T}\widehat{\Delta}_{1t}^{(\alpha)}\nonumber  &
	=\frac{1}{T_2}\sum_{t=T_1+1}^T\left(y_{1t}-\widehat{\delta}_{1,\alpha}-\boldsymbol{\widehat{\delta'}_\alpha\widetilde{y_t}}\right)\nonumber\\  
	&=\int\left(y_1-\boldsymbol{x}'\boldsymbol{U}_{\alpha}\left(G_{n,pre}\right)\right)dG_{n,post}(\boldsymbol{x},y_{1}),
\end{align}
where $\boldsymbol{U}_{\alpha}(.)$ denotes the functional for the MDPDE $\boldsymbol{\widehat{\beta}}_\alpha=(\widehat{\delta}_{1,\alpha},\boldsymbol{\widehat{\delta}_\alpha}).$ 
Therefore, the statistical functional corresponding to the Mean-MDPDE of the ATE would be defined as 
\begin{equation}\label{6.3}
	T_{\alpha}^\ast\left(G_{pre},G_{post}\right) =\int\left(y_1-\boldsymbol{x}'\boldsymbol{U}_{\alpha}\left(G_{pre}\right)\right)dG_{post}(\boldsymbol{x},y_{1}).
\end{equation}

To derive the influence function of this ATE functional, we consider the pre- and post-treatment contaminated distributions given by  
$$
G_{pre,\epsilon}=(1-\epsilon)G_{pre}+\epsilon \wedge(\boldsymbol{x}_t,y_{1t})
~~~~\mbox{and}~~ 
G_{post,\epsilon}=(1-\epsilon)G_{post}+\epsilon \wedge(\boldsymbol{x}_t,y_{1t}),
$$
where $\epsilon$ is the contamination proportion and  
$\wedge(.,.)$ denotes the distribution function for a degenerate distribution at the point of contamination $(\boldsymbol{x}_t,y_{1t})$.
We will then derive the influence functions for the case of pre-treatment contamination and that of the post-treatment contamination in the following subsections.

\subsection{Influence function for Pre-Treatment Contamination}

In order to derive the influence function of the Mean-MDPDE ATE functional $T_{\alpha}^\ast\left(G_{pre},G_{post}\right)$, 
we need to substitute $G_{pre}$ by $G_{pre,\epsilon}$ in \eqref{6.3} and differentiate it with respect to $\epsilon$. 
Evaluating the result at $\epsilon=0$, we get the desired influence function as
\begin{align}
IF_0((\boldsymbol{x}_t,y_{1t}), T_{\alpha}^\ast, (G_{pre},G_{post})) 
&= \left. \frac{\partial}{\partial \epsilon} T_{\alpha}^\ast\left(G_{pre,\epsilon},	G_{post}\right)\right|_{\epsilon=0} \nonumber\\
&= \int\left(\boldsymbol{x}'\left.\frac{\partial}{\partial\epsilon}\boldsymbol{U}_{\alpha}\left(G_{pre,\epsilon}\right)\right|_{\epsilon=0}\right)dG_{post} \nonumber\\
	&=\int\boldsymbol{x}'IF\left((\boldsymbol{x}_t,y_{1t});\boldsymbol{U}_{\alpha},G_{pre}\right)dG_{post}.
\end{align}
But, it can be easily shown that the influence function for the MDPDE functional $\boldsymbol{U}_{\alpha}$ at the normal model density is given by 
(see \cite{Basu/etc:2011,Ghosh/Basu:2013})
$$
IF(\boldsymbol{x}_t,y_{1t});T,G_{pre})=(1+\alpha)^{3/2}E(\boldsymbol{X}_0'\boldsymbol{X}_0)^{-1}\boldsymbol{x}_t(y_{1t}-\beta'\boldsymbol{x}_t)
e^{-\frac{\alpha(y_{1t}-\boldsymbol{\beta}'\boldsymbol{x}_t)^2}{2\sigma^2}}. 
$$
Therefore, the influence function for the Mean-MDPDE of the ATE under pre-treatment contamination turns out to be  
$$
IF_0((\boldsymbol{x}_t,y_{1t}), T_{\alpha}^\ast, (G_{pre},G_{post})) =
-(1+\alpha)^{3/2}(y_{1t}-\boldsymbol{\beta}'\boldsymbol{x}_t)e^{-\frac{\alpha(y_{1t}-\beta'\boldsymbol{x}_t)^2}{2\sigma^2}}
\int \boldsymbol{x}'E(\boldsymbol{X}_0'\boldsymbol{X}_0)^{-1}\boldsymbol{x}_t\hspace{1mm}dG_{post}(\boldsymbol{x}).
$$
It is easy to see that the above influence function is bounded with respect to contamination ($y_{1t}$) in the response variable for all $\alpha>0$, 
justifying our empirical findings about the robustness of the Mean-MDPDE of the ATE. 
However, its boundedness, and hence the robustness, with respect to contamination in covariate space depends on the distribution of the covariates.

As a special case, the above influence function  also provides the same for the classical HCW estimator of the ATE at $\alpha=0$,
which turns out to be 
$$
IF_0((\boldsymbol{x}_t,y_{1t}), T_{0}^\ast, (G_{pre},G_{post})) =
-(y_{1t}-\boldsymbol{\beta}'\boldsymbol{x}_t)\int \boldsymbol{x}'E(\boldsymbol{X}_0'\boldsymbol{X}_0)^{-1}\boldsymbol{x}_t\hspace{1mm}dG_{post}(\boldsymbol{x}).
$$
This function is clearly unbounded even in $y_{1t}$ indicating the non-robust nature of the HCW estimator even under infinitesimal contamination. 

\begin{figure}[h]
	\centering
	\includegraphics[width=0.7\textwidth]{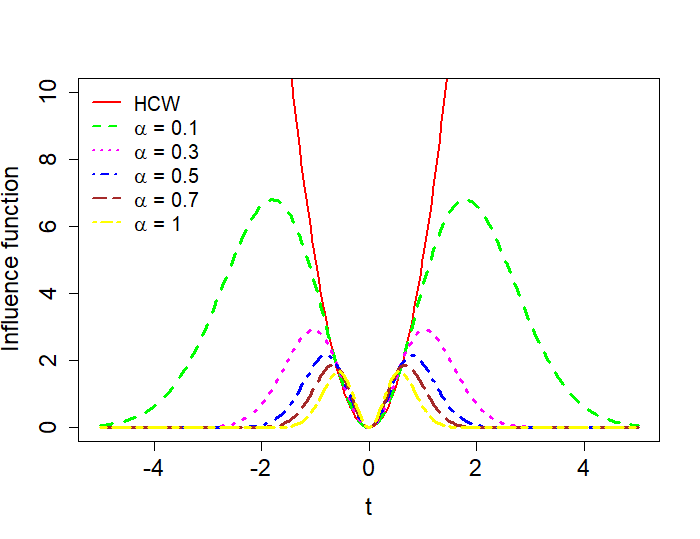}
	\caption{Influence function of the Mean-MDPDE of the ATE under pre-treatment contamination}
	\label{FIG:IF}
\end{figure}

\bigskip
\noindent\textbf{An Example:}\\ 
For illustration, we plot an example of the influence function of the Mean-MDPDE of the ATE for different $\alpha>0$,
and the same for the HCW estimator, in Figure \ref{FIG:IF}.
Here, we have considered $\boldsymbol{x}=(1,y_2)'$, $\boldsymbol{x}_t=(t,t)$, $y_{1t}=t$ and $\delta_1=2$, $\delta=\beta_2=4$, $\sigma=2$.
Clearly, the influence function of the classical HCW estimator is unbounded while we get a bounded influence function for the robust Mean-MDPDE 
at any tuning parameter $\alpha>0$. Further, the maximum value of these influence functions decreases with increasing values of $\alpha>0$ 
indicating the increase in the extent of robustness of the corresponding Mean-MDPDE of the ATE.

\subsection{Influence function for Post-Treatment Contamination}

For contamination in the post-treatment distribution, we substitute $G_{post,\epsilon}$ in place of $G_{post}$ in \eqref{6.3} and 
again differentiate with respect to $\epsilon$ at $\epsilon=0$. 
This leads to the influence function of the  Mean-MDPDE of the ATE under post-treatment contamination as given by 
%\begin{align}
%	T_{\alpha}^\ast\left(G_{pre,},G_{post,\epsilon}\right) &=\int\left(y_1-\boldsymbol{x}'\boldsymbol{U}_{\alpha}\left(G_{pre}\right)\right)dG_{post,\epsilon} \nonumber\\
%	&=\int\left(y_1-\boldsymbol{x}'\boldsymbol{U}_{\alpha}\left(G_{pre}\right)\right)d\left(1-\epsilon\right)G_{post}+\epsilon D\left(\boldsymbol{x}_t,y_{1t}\right)
%\end{align}
%Taking derivative with respect to $\epsilon$ and putting $\epsilon=0$, we get 
\begin{align*}
IF_1((\boldsymbol{x}_t,y_{1t}), T_{\alpha}^\ast, (G_{pre},G_{post})) 
&=	\left.\frac{\partial}{\partial \epsilon} T_{\alpha}^\ast\left(G_{pre},	G_{post,\epsilon}\right)\right|_{\epsilon=0} \nonumber\\
&= \int\left(y-\boldsymbol{x}'\boldsymbol{U}_{\alpha}\left(G_{pre}\right)\right)d[\wedge\left(\boldsymbol{x}_t,y_{1t}\right)-G_{post}]\\ 
&= y_{1t}-\boldsymbol{x}_t'\boldsymbol{U}_{\alpha}\left(G_{pre}\right)-\int\left(y_1-\boldsymbol{x}'\boldsymbol{U}_{\alpha}\left(G_{pre}\right)\right)dG_{post}\\ 
&= y_{1t}-\boldsymbol{x}_t'\boldsymbol{U}_{\alpha}\left(G_{pre}\right)-T_{\alpha}^\ast\left(G_{pre}, G_{post}\right).
\end{align*}
Note that, this influence function is a linear function in the contamination points $\boldsymbol{x}_t$ and $y_{1t}$, and hence unbounded for any $\alpha\geq 0$. 
This further validates the results of our simulation study indicating that the Mean-MDPDE of the ATE is not suitable for the cases 
where we have contamination in the post treatment period.

\bigskip
\noindent\textbf{An Alternative Estimator for Post-Treatment Contamination:}\\ 
%\subsection{An Alternative Estimator for Post-Treatment Data Contamination}\label{6.03}
Note that the main reason for the non-robustness of the Mean-MDPDE of the ATE under contamination in post-treatment period is the use of mean
to get a summary measures from the individual ATE estimators $\widehat{\Delta}_{1t}^{(\alpha)}$ at the post-treatment period via Equation (\ref{EQ:Mean-MDPDE}),
and mean is known to be a non-robust summary measure.
So, in case of post-treatment contamination, we may get a robust summary estimate of the ATE by considering the median of $\widehat{\Delta}_{1t}^{(\alpha)}$s. 
This estimator, which we will refer to as the Median-MDPDE of the ATE, is formally defined as 
\begin{equation}\label{6.7}
%	\widetilde{\Delta}_1^{(\alpha)} = \mbox{Median}_{T_1+1\leq t\leq T}\widehat{\Delta}_{1t}^{(\alpha)}.
	\widetilde{\Delta}_1^{(\alpha)} = \mbox{Median}\left\{\widehat{\Delta}_{1t}^{(\alpha)} ~:~ T_1+1\leq t\leq T\right\}.
\end{equation}
To illustrate the claimed robustness of this Median-MDPDE of the ATE, 
we repeat the simulation study performed in Section \ref{SEC:SIMULATION} to compare the empirical bias and MSE of $\widetilde{\Delta}_1^{(\alpha)}$.
The results obtained for the post-treatment contamination are reported in Tables \ref{TAB:Bias2} and \ref{TAB:MSE2};
results for the other cases are very similar to those obtained for the Mean-MDPDE of the ATE and hence they are not reported again for brevity. 
From these tables, it is evident that the Median-MDPDE of the ATE can successfully produce robust estimators in the presence of data contamination 
in post-treatment period in terms leading to significantly smaller MSE and bias compared to the Mean-MDPDE and all other existing ATE estimators considered here.

\begin{table}[!h]
	\caption{Empirical biases of the Median-MDPDE of ATE at different $\alpha> 0$} 
	\centering
	\begin{adjustbox}{width=.9\textwidth}
		\begin{tabular}{|l|ccc|ccc|ccc|}
			\hline
			$T_1$&\multicolumn{3}{|c|}{100}&\multicolumn{3}{|c|}{200}&\multicolumn{3}{|c|}{400}\\
			%			\hline
			$T_2$&20&80&320&20&80&320&20&80&320\\
			\hline
			%			\multicolumn{10}{|c|}{$5\%$ of post treatment $u_{1t}\sim N(5,1)$, remaining $u_{jt}\sim N(0,1)$}\\
			%\hline
			\multicolumn{10}{|c|}{\textbf{$5\%$ Contamination in post treatment period}}\\						
%			$\alpha=0$&0.0765&0.0834&0.0767&0.0818&0.08&0.0777&0.0852&0.0818&0.0754\\
			$\alpha$=0.1 &0.0766&0.0836&0.0763&0.0819&0.0795&0.0776&0.0853&0.0819&0.0754\\
			$\alpha=0.3$&0.0765&0.0833&0.0762&0.082&0.0788&0.0777&0.0853&0.0819&0.0751\\
			$\alpha=0.5$&0.0762&0.083&0.0754&0.082&0.0786&0.0777&0.0852&0.082&0.0748\\
			$\alpha=0.7$&0.076&0.0824&0.0754&0.082&0.0783&0.0776&0.0851&0.0819&0.0748\\
			$\alpha=1$&0.0747&0.0824&0.0757&0.082&0.0779&0.0775&0.0849&0.082&0.075\\
			\hline
			%			\multicolumn{10}{|c|}{$20\%$ of post treatment $u_{1t}\sim N(5,1)$, remaining $u_{jt}\sim N(0,1)$}\\
			%			\hline
			\multicolumn{10}{|c|}{\textbf{$20\%$ Contamination in post treatment period}}\\						
%			$\alpha=0$&0.4015&0.3862&0.3765&0.4132&0.3871&0.3801&0.4196&0.3887&0.3771\\
			$\alpha=0.1$&0.4015&0.3865&0.3762&0.413&0.3867&0.38&0.4198&0.389&0.377\\
			$\alpha=0.3$&0.4015&0.387&0.3761&0.4127&0.386&0.38&0.42&0.3893&0.3766\\
			$\alpha=0.5$&0.4014&0.3874&0.3761&0.4126&0.3856&0.3802&0.4202&0.3896&0.3765\\
			$\alpha=0.7$&0.4017&0.3876&0.3766&0.4127&0.3852&0.3806&0.4203&0.3898&0.3765\\
			$\alpha=1$&0.4021&0.388&0.3774&0.4131&0.3851&0.3811&0.4205&0.3902&0.3769\\
			\hline
		\end{tabular}
	\end{adjustbox}
	\label{TAB:Bias2}
\end{table}

\begin{table}[!h]
	\caption{Empirical MSEs of the Median-MDPDE of ATE at different $\alpha> 0$} 
	\centering
	\begin{adjustbox}{width=.9\textwidth}
		\begin{tabular}{|l|ccc|ccc|ccc|}
			\hline
			%			\multicolumn{10}{|c|}{$5\%$ of post treatment $u_{1t}\sim N(5,1)$, remaining $u_{jt}\sim N(0,1)$}\\
			%			\hline
			$T_1$&\multicolumn{3}{|c|}{100}&\multicolumn{3}{|c|}{200}&\multicolumn{3}{|c|}{400}\\
			%			\hline
			$T_2$&20&80&320&20&80&320&20&80&320\\
			\hline
			\multicolumn{10}{|c|}{\textbf{$5\%$ Contamination in post treatment period}}\\						
%			$\alpha=0$&0.14&0.057&0.0296&0.1409&0.0443&0.0219&0.1424&0.0425&0.0178\\
			$\alpha$=0.1 &0.1403&0.0577&0.0297&0.1412&0.0444&.0219&0.1424&0.0426&0.0178\\
			$\alpha=0.3$&0.1414&0.0594&0.031&0.1423&0.0449&0.0223&0.1427&0.0429&0.0179\\
			$\alpha=0.5$&0.1436&0.0614&0.0326&0.1435&0.0458&0.0229&0.1432&0.0432&0.0182\\
			$\alpha=0.7$&0.1467&0.0636&0.0349&0.145&0.047&0.0236&0.1438&0.0435&0.0187\\
			$\alpha=1$&0.1515&0.0673&0.0389&0.1474&0.0486&0.0251&0.145&0.0442&0.0196\\
			\hline
			%			\multicolumn{10}{|c|}{$20\%$ of post treatment $u_{1t}\sim N(5,1)$, remaining $u_{jt}\sim N(0,1)$}\\
			%			\hline
			\multicolumn{10}{|c|}{\textbf{$20\%$ Contamination in post treatment period}}\\						
%			$\alpha=0$&0.3872&0.2169&0.1697&0.4013&0.2061&0.1655&0.4064&0.2043&0.1585\\
			$\alpha=0.1$&0.3874&0.2172&0.1697&0.4013&0.2061&0.1655&0.4064&0.2044&0.1584\\
			$\alpha=0.3$&0.3889&0.2187&0.1709&0.4019&0.2064&0.1658&0.4067&0.2049&0.1584\\
			$\alpha=0.5$&0.3914&0.2206&0.1728&0.4029&0.2068&0.1666&0.407&0.2054&0.1587\\
			$\alpha=0.7$&0.395&0.2229&0.1754&0.4043&0.2075&0.1677&0.4074&0.2061&0.1591\\
			$\alpha=1$&0.4003&0.2274&0.1802&0.4068&0.209&0.1694&0.4084&0.2072&0.1602\\
			\hline
		\end{tabular}
	\end{adjustbox}
	\label{TAB:MSE2}
\end{table}

\section{Real Data Example: Long-term Effects of Tsunami on GDP }
\label{SEC:data}

We will now illustrate the applicability of the proposed ATE estimator in a practical scenario of investigating the long term effects of 
the ``\textit{2004 Indian Ocean Tsunami}" \cite{Tsunami} on the per capita GDP of the most severely affected countries. 
The tsunami of 2004 that originated in the Indian Ocean severely affected the surrounding countries causing damage to innumerable lives and properties. 
We want to estimate the effects of this natural calamity on the per capita GDP of the five severely hit countries 
which are Indonesia, India, Thailand, Maldives and Sri Lanka. 
For this purpose, we collect the data on per capita GDP (in US \$) from \url{https://www.macrotrends.net/countries/ranking/gdp-per capita}, 
over period 1981--2004 as the pre-treatment period and from 2005 to 2019 as the post-treatment period. 
As for the control group, it is natural to consider the same data for the countries which were not affected by the hazard. 
For our study, the countries in the control group are Portugal, Iceland, Israel, Netherlands, Brazil, USA, Japan, Ethiopia, 
Switzerland, Tunisia, Bangladesh, Pakistan, Philippines and Egypt, covering a wide range of economic conditions.

Based on some preliminary analyses, the logarithm of the per capita GDP of countries in the control group is taken as the covariate 
while the response is considered to be the logarithm of the per capita GDP of each of the affected (treated) countries. 
Then, we estimate the ATE of the Tsunami based on these data via the proposed Mean-MDPDE, Median-MDPDE and also using other existing ATE estimators;
the resulting ATE estimates are reported in Table \ref{TAB:data_ATE}. 
Note that, these results can be viewed as a long-term effects of Tsunami since they are computed summarizing the effects from 5 years in the post-treatment period. 
{\cbl The values of optimum $\alpha$ obtained by the moving block bootstrap method, as described in Section \ref{SEC:Sel_alpha},
	and the corresponding Mean and Median-MDPDEs are also reported in the same Table \ref{TAB:data_ATE}. }

\begin{table}[!h]
	\centering
	\caption{Different Estimators of the (long-term) ATE of Tsunami on per capita GDP of five countries based on data 
		from 1981--2004 (pre-treatment) and 2005--2009 (post-treatment)}
	\color{black}
%\resizebox{.92\textwidth}{!}{
	\begin{tabular}{|l|r|r|r|r|r|}
		\hline
		Estimators & Indonesia & India & Thailand & Maldives & Sri Lanka\\
		\hline
		HCW &$-$1.04&$-$0.064&$-$0.289&0.311&0.784\\
		DID&$-$0.518&$-$0.404&$-$0.307&$-$0.914&$-$0.689\\
		ADID &0.736&0.741&$-$0.180&$-$0.331&0.429\\
		MSCM &0.754&0.705&0.460&1.094&0.945\\
	\hline
	Mean-MDPDE($\alpha=0.1$) & -0.918  &  -0.131 & -0.299 & 0.093 & 0.844\\ 
	Mean-MDPDE($\alpha=0.3$) & -1.239  & -0.081 & -0.528 & -0.003 & 0.636\\ 
	Mean-MDPDE($\alpha=0.5$) & -0.921 & 0.218 & -0.368 & -0.161 & 0.908\\ 
	Mean-MDPDE($\alpha=0.7$) & -0.798  & -0.336 &-0.435 & -0.048& 1.018\\ 
%	($\alpha_{opt}$) & (0.45)  &  (0.40) &  (0.1) &  (0.25) & (0.65)\\ 
		\hline
Median-MDPDE($\alpha=0.1$) & -1.101  &  -0.15 & -0.391 & 0.079 & 0.786\\ 
Median-MDPDE($\alpha=0.3$) & -1.406  & -0.098 & -0.659 & -0.049 & 0.523\\ 
Median-MDPDE($\alpha=0.5$) & -1.103   &  0.228 & -0.445 & -0.159 & 0.838\\ 
Median-MDPDE($\alpha=0.7$) & -0.927  &  -0.356 & -0.541 & -0.104 & 0.887\\
\hline
$\alpha_{opt}$ &  (0.45)  &  (0.40) &  (0.10) & (0.25) & (0.65)\\ 
Mean-MDPDE($\alpha_{opt}$) & -0.805 &  0.008 & -0.299 & 0.004 & 0.941 \\ 
Median-MDPDE($\alpha_{opt}$) & -0.828   &  -0.031 & -0.391  & -0.060  & 0.824 \\ 
		\hline		
	\end{tabular}
%}
\label{TAB:data_ATE}
\end{table}

%\begin{figure}[!h]
%	\centering
%	\subfloat[Indonesia]{\includegraphics[width=.5\linewidth]{indonesia.png}}
%	\subfloat[India]{\includegraphics[width=.5\linewidth]{india.png}}
%	\\ %--------------------------------------------------------------------
%	\subfloat[Thailand]{\includegraphics[width=.5\linewidth]{thailand.png}}
%	\subfloat[Maldives]{\includegraphics[width=.5\linewidth]{maldives.png}}
%	\\ %--------------------------------------------------------------------
%	\subfloat[Sri Lanka]{\includegraphics[width=.5\linewidth]{srilanka.png}}
%%	\subfloat[India]{\includegraphics[width=.5\linewidth]{india.png}}
%%---------------------------------------------------------------------------
%	\caption{Observed data-points and the fitted curves based on the HCW estimator and the proposed Mean-MDPDE for several $\alpha>0$.}
%\label{FIG:data}
%\end{figure}

{\cbl Here we should point out that the HCW method required the data to be stationary which is  not the case here 
	with the logarithmic GDP values considered, but we have still kept this estimate among our results as a benchmark for comparison.
	On the contrary, our proposed estimators can indeed handle the non-stationary data and so are more appropriate for such applications. 
It can be seen from Table \ref{TAB:data_ATE} that our proposal gives a valid way to compute the average treatment effect 
which is comparable with other existing estimates in cases with no contamination  
and provide a better robust estimates in cases with possible data contamination.}
That the proposed procedures give a better fit to the data with all moderate and optimum $\alpha$ values has  been confirmed by looking at the corresponding coefficient of determination ($R^2$), which are not reported here for brevity. 
%These results, in turn, imply that the ATE estimates obtained by the proposed Mean-MDPDE (or the Median-MDPDE) are more accurate 
%as far as the fit to the overall observed data is concerned and, hence, they are able to provide correct (robust) insights 
%in this particular application where existing methods have failed drastically. 

\begin{figure}[!h]
	\centering
	\includegraphics[width=0.78\linewidth]{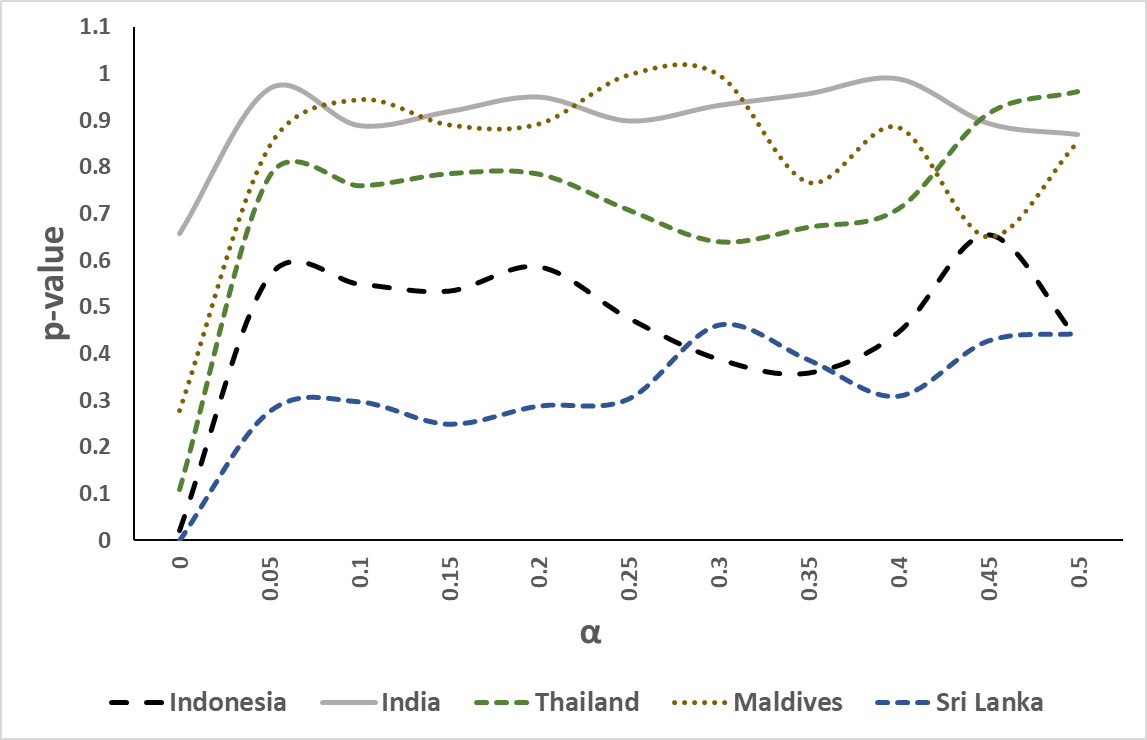}
	\caption{p-values for testing of significance of the ATEs  by the Wald-type tests using the HCW estimator (at $\alpha=0$)
			and the proposed Mean-MDPDE of the ATE for several $\alpha>0$. }
	\label{FIG:data_p-value}
\end{figure}

To investigate further, we note that the estimated ATE based on the Mean-MDPDE (or the Median-MDPDE) with $\alpha>0$ 
are quite close to zero. So, we next test for the significance of these estimated ATEs 
by testing the hypothesis $H_0 : \Delta_1 = 0$ against $H_1 : \Delta_1 \neq 0$. 
We perform this test based on the proposed MDPDE-based Wald-type test statistics to get robust insights;
the resulting $p$-values are plotted over the tuning parameter $\alpha\geq 0$ in Figure \ref{FIG:data_p-value}.
Note that, the p-value at $\alpha=0$ corresponds to the test of significance based on the existing HCW estimator. 
It is again evident that the null hypothesis of no significant effect is rejected for Indonesia and Sri Lanka at 5\% level of significance 
(p-value for Thailand is also quite low as 0.108) while using the existing HCW estimator, 
but the proposed MDPDE based Wald-type tests, performed using the Mean-MDPDE of the ATE, 
give much more stable and high p-values for all the five countries. 
Based on these robust inference, we can then conclude that there is no significant average effect of tsunami on the per capita GDP 
of each of these countries over the period from 2005-2019.

It is important to note here that the above conclusion, although supported by the data, may appear to be non-intuitive at a first glance 
since such natural disasters are deemed to have negative economic consequences. 
However, the consequences of the 2004 tsunami were quite unexpected and so the subsequent international humanitarian responses 
towards the tsunami affected countries were swift and outstanding.
Athukorala \cite{Athukorala:2012} wrote that it was the largest ever international aid commitment in terms of both the total volume of trade and aid per affected person. 
It also helped people of these developing countries not to suffer from diseases and starvation that usually follow such disasters. 
All these made it possible for these Tsunami-affected countries to recover promptly without having any long-term impact in their per capita GDP. 
In fact, it has also been argued that \cite{Heger/Neumayer:2019} such copious influx of international assistance triggered higher economic growth in the Aceh province of Indonesia than what would have happened in the absence of Tsunami.
The same is also true for the other affected countries as supported by our empirical investigation based on the proposed robust procedures,
{\cbl but it definitely needs further economic investigation to confirm this empirical insights 
	as there can indeed be many other hidden factors.  
But, this example clearly illustrates the usefulness of our proposed methodology in real data analyses 
which has been the main purpose of this paper.}

\section{Concluding Remarks}
\label{SEC:conclusion}

In real-life data, we often encounter unusual observations (outliers) that make  analyses cumbersome. 
They often mislead the path of analysis if not treated properly. In order to circumvent their influence, 
we prefer robust methods that are not affected by the presence of data contamination. 
In the light of this need, we have developed a robust estimator of the ATE from panel data based on the popular approach of robust  MDPDE. 
A rigorous study of the large sample properties of the new ATE estimator is carried out. 
Statistical tests of hypotheses based on the proposed ATE estimator are also developed. 
Theoretical study of the robustness of the proposed estimator is manifested through the influence function analysis. 
The real life  performance of the new estimator is exemplified through the analysis of the effect of tsunami of 2004 on the per capita GDP of few severely hit countries 
where the traditional method failed to perform well in the post-treatment period but our proposed procedures have exhibited satisfactory results.

As mentioned before, we hope to study the data-driven algorithms for choosing an optimum $\alpha$ values for a given dataset 
in more details for complex contexts in our future research. 
We also hope to extend the proposed methodology for robust estimation of the ATE from other types of data in our sequel research works.

\bigskip
\noindent\textbf{Acknowledgment:}\\
This research is partially supported by an internal research grant from Indian Statistical Institute, India. 
Research of AG is also partially supported by an INSPIRE Faculty Research Grant from Department of Science and Technology, Government of India.

%\newpage
\appendix
\section{Brief Background on the Minimum DPD Estimation}
\label{APP:MDPDE}

%\subsection{Minimum Density Power Divergence Estimator}
%\renewcommand{\theequation}{A.\arabic{equation}}

The DPD is a measure of discrepancy between two probability density functions, 
proposed for the robust estimation of model parameters by \citet{Basu/etc:1998}. 
Given two densities $g$ and $f$, the DPD measure is defined in terms of a single tuning parameter $\alpha$ as
\begin{equation}\label{2.7}
	d_\alpha(g,f) = \int\left\{f^{1+\alpha}(\boldsymbol{x})-\left(1+\frac{1}{\alpha}\right)g(\boldsymbol{x})f^\alpha(\boldsymbol
	{x})+\frac{1}{\alpha}g^{1+\alpha}(\boldsymbol{x})\right\}d\boldsymbol{x}, ~~~~\alpha>0.
\end{equation}
At $\alpha=0$, the divergence is expressed as the limiting form:
\begin{align}
	d_0(g,f) = \lim_{\alpha\to 0} d_\alpha(g,f) = \int g(\boldsymbol{x})\ln\left(\frac{g(\boldsymbol{x})}{f(\boldsymbol{x})}\right)d\boldsymbol{x}.
\end{align}
which is the Kullback-Leibler divergence (KLD) leading to the most efficient (extremely non-robust) MLE. 
When $\alpha=1$, the divergence is simply the integrated squared error and the parameter estimates are highly robust (although less efficient). 
Thus, the tuning parameter $\alpha$ controls the trade-off between robustness and asymptotic efficiency of the resulting parameter estimator;
the degree of robustness increases but the (pure data) efficiency decreases with an increase in $\alpha$.

Now, let us first consider the independent and identically distributed (IID) observations $X_1, X_2,...X_n$ from a population
having true distribution function $G$ and let $g$ denote the corresponding true density function. 
We model $g$ by a parametric family of densities $\mathcal{F}=\{f_{\boldsymbol{\theta}}(x): \boldsymbol{\theta}\in\Theta\subseteq \mathbb{R}^p\}$. 
Then, the minimum DPD functional  at the distribution $G$ is defined as the minimizer of $d_\alpha(g,f_{\boldsymbol{\theta}})$ over $\boldsymbol{\theta} \in \Theta$.
In practice, the corresponding minimum DPD estimator (MDPDE) is defined as the minimum DPD functional evaluated at the empirical distribution function $G_n$ 
obtained based on the sample observations. Note that the third term of the divergence function $d_\alpha(g,f) $ in \eqref{2.7} will not play any role 
in the minimization process, and hence can be dropped from the objective function.  So, rewriting the integral in the second term in DPD as
$\int g(\boldsymbol{x})f^\alpha(\boldsymbol{x})d\boldsymbol{x} = \int f^\alpha(\boldsymbol{x})dG(\boldsymbol{x})$ and subsequently using $G_n$ in place of $G$,
we obtain the MDPDE  $\widehat{\boldsymbol{\theta}}$ of $\boldsymbol{\theta}$ by minimizing the resultant objective function  \cite{Basu/etc:1998}
\begin{equation}
	H_n(\boldsymbol{\theta}) = \int f_{\boldsymbol{\theta}}^{1+\alpha}(x)dx-\left(1+\frac{1}{\alpha}\right)\frac{1}{n}\sum_{i=1}^{n}f_{\boldsymbol{\theta}}^\alpha(X_i),
	\label{EQ:MDPDE}
\end{equation}
with respect to $\boldsymbol{\theta}\in\Theta$, for any $\alpha>0$. 
Assuming differentiability of the model densities and putting $\boldsymbol{u_\theta}= \frac{\partial}{\partial\boldsymbol{\theta}}\ln f_{\boldsymbol{\theta}}$, 
the MDPDE can equivalently be obtained by solving the estimating equations given by
\begin{equation}
	\frac{1}{n}\sum_{i=1}^{n}\boldsymbol{u_\theta}(X_i)f_{\boldsymbol{\theta}}^\alpha(X_i)
	-\int \boldsymbol{u_\theta(x)}f_{\boldsymbol{\theta}}^{1+\alpha}(\boldsymbol{x})dx=0, ~~~~ \alpha\geq0.
\end{equation}

Later, \citet{Ghosh/Basu:2013} generalized the concept of robust minimum DPD estimation to the cases of independent but not identically distributed observations 
that arise, e.g., in fixed-design parametric regressions. 
Suppose that the observed data $Y_1, Y_2,\cdots,Y_n$ are independent but, for each $i$, $Y_i\sim g_i$ 
with $g_1, \cdots, g_n$ being some possibly different densities with respect to some common dominating measure. 
To model $g_i$'s, we use the parametric family $\mathcal{F}_i=\{f_i(.;\boldsymbol{\theta}): \boldsymbol{\theta}\in\Theta\}$ for $i=1,2,\ldots$. 
In this case, although the model densities are different, they share the same parameter $\boldsymbol{\theta}$. 
\citet{Ghosh/Basu:2013} proposed to compute the DPD measure $d_{\alpha}(g_i,f_i(.;\boldsymbol{\theta})$ between data and model separately  for each data point 
and then minimize their average divergence, i.e., $\dfrac{1}{n}\sum\limits_{i=1}^{n}d_{\alpha}(g_i,f_i(.;\boldsymbol{\theta}))$ with respect to 
$\boldsymbol{\theta} \in \Theta$. 
%Noting that the third term in DPD does not involve the parameter, \citet{Ghosh/Basu:2013} simplified 
As before, the objective function for the MDPDE can again be simplified, in such non-homogeneous cases, to have the form 
\begin{align}
H_n(\boldsymbol{\theta}) = \dfrac{1}{n}\sum_{i=1}^{n}\left[\int f_i(y,\boldsymbol{\theta})^{1+\alpha}dy 
-\left(1+\frac{1}{\alpha}\right)f_i(Y_i,\boldsymbol{\theta})^\alpha\right].
\label{EQ:MDPDE_INH}
\end{align}
Differentiating the above with respect to $\boldsymbol{\theta}$, we get the MDPDE estimating equations as given by 
\begin{equation}
\sum_{i=1}^{n}\left[f_i(Y_i;\boldsymbol{\theta})^{\alpha}\boldsymbol{u}_i(Y_i,\boldsymbol{\theta}) 
- \int f_i(y,\theta)^{1+\alpha}u_i(y,\theta)dy\right]=0.
\label{EQ:MDPDE_INHe}
\end{equation}
The MDPDEs are needed to obtain numerically by minimize the objective function (\ref{EQ:MDPDE_INH}) or solving the estimating equation (\ref{EQ:MDPDE_INHe}).
The asymptotic and robustness properties of the MDPDE under such general non-homogeneous set-ups are derived in \cite{Ghosh/Basu:2013}.
These ideas are also applied to study robust MDPDE for the linear regression model in \cite{Ghosh/Basu:2013} and 
for the more general class of generalized linear models in \cite{Ghosh/Basu:2016b,Ghosh:2019}.

\section{Proof of Theorem \ref{PROP:Asymp_MDPDE}}
\label{APP:Proof_MDPDEs}

We first note that our set-up is exactly similar to the set-up of independent non-homogeneous observations, as described in Appendix \ref{APP:MDPDE},
with the individual densities being given by (under the notations of Section \ref{SEC:Asymp_MDPDE})
\begin{align}
&f_t(y_{1t})=\frac{1}{\sigma}f\left(\frac{y_{1t}-\delta_1-\boldsymbol{\delta'\widetilde{ y_t}}}{\sigma}\right)
=\frac{1}{\sigma}f\left(\frac{y_{1t}-\boldsymbol{\beta' { x_t}}}{\sigma}\right), ~~~~~t=1, \ldots, T_1.
\label{3.1}
%\nonumber\\
%	&\therefore ln(f_t(y_{1t}))=-ln \sigma+ln\left(f\left(\frac{y_{1t}-\boldsymbol{\beta' { x_t}}}{\sigma}\right)\right)\nonumber\\
%	&\hspace{2.2cm}=-\frac{1}{2}ln(\sigma^2)+ln\left(f\left(\frac{y_{1t}-\boldsymbol{\beta' { x_t}}}{\sigma}\right)\right)
\end{align}
Therefore, the existence and the asymptotic properties of the MDPDE follows directly from Theorem 3.1 of \cite{Ghosh/Basu:2013}
with the asymptotic variance being given by $\boldsymbol{\Psi}^{-1}\boldsymbol{\Omega}\boldsymbol{\Psi}^{-1}$, where 
$$
\boldsymbol{\Psi}_\alpha=\frac{1}{T_1}\sum_{t=1}^{T_1}\boldsymbol{J}^{(t)}, ~~\mbox{ and  }~~~
\boldsymbol{\Omega}_\alpha=\frac{1}{T_1}\sum_{t=1}^{T_1}\bigg[\int \boldsymbol{u}_t(y_1)\boldsymbol{u}_t(y_1)'f_t^{2\alpha+1}(y_1)dy_1-\xi\xi'\bigg],
$$
with  $\boldsymbol{u}_t(y_1) = \frac{\partial }{\partial(\boldsymbol{\beta}', \sigma)'}\ln f_t(y_1)$, 
$\boldsymbol{J}^{(t)}=\int \boldsymbol{u}_t(y_1)\boldsymbol{u}_t(y_1)'f_t^{\alpha+1}(y_1)dy_1$ and  
$\boldsymbol{\xi}_t=\int \boldsymbol{u}_t(y_1) f_t^{\alpha+1}(y_1)dy_1$.
Thus, the proof of the present theorem will be completed once we simplify the matrices $\boldsymbol{\Psi}_\alpha$ and $\boldsymbol{\Omega}_\alpha$
under our particular model set-up and assumptions. 

Now, taking logarithm of $f_t(y_{1t})$ in \eqref{3.1} and differentiating it with respect to  $\boldsymbol{\beta}$ and $\sigma^2$, we get
\begin{align}
&\frac{\partial ln(f_t(y_{1t}))}{\partial \beta}=\frac{f'\left(\frac{y_{1t}-\boldsymbol{\beta' x_t}}{\sigma}\right)}{f\left(\frac{y_{1t}-\boldsymbol{\beta' x_t}}{\sigma}\right)}\times \frac{-\boldsymbol{x_t}}{\sigma},\nonumber\\
&\frac{\partial ln(f_t(y_{1t}))}{\partial \sigma^2}=-\frac{1}{2\sigma^2}-\frac{f'\left(\frac{y_{1t}-\boldsymbol{\beta' x_t}}{\sigma}\right)}{f\left(\frac{y_{1t}-\boldsymbol{\beta' x_t}}{\sigma}\right)}\times \left(\frac{y_{1t}-\boldsymbol{\beta'x_t}}{2\sigma^3}\right),\nonumber
\end{align}
and hence 
$
\boldsymbol{u}_t\left(y_{1t}\right)= - \left(\begin{array}{c}
	u\left(\frac{y_{1t}-\boldsymbol{\beta' x_t}}{\sigma}\right)\left(\frac{\boldsymbol{x_t}}{\sigma}\right)\\
	\frac{1}{2\sigma^2}+u\left(\frac{y_{1t}-\boldsymbol{\beta' x_t}}{\sigma}\right) \left(\frac{y_{1t}-\boldsymbol{\beta'x_t}}{2\sigma^3}\right)
\end{array}\right)$.

\noindent
Then, we calculate the matrix $\boldsymbol{J}^{(t)}$ via the partition 
$\boldsymbol{J}^{(t)}=\begin{bmatrix}
	A&B\\
	B'&C\\
\end{bmatrix}$, 
where 
\begin{eqnarray}
A 	&=& \frac{\boldsymbol{x}_t \boldsymbol{x}_t' }{\sigma^{\alpha+3}}\int\bigg\{u\left(\frac{y_{1t}-\boldsymbol{\beta'x_t}}{\sigma}\right)\bigg\}^2 
		f\left(\frac{y_{1t}-\boldsymbol{\beta'x_t}}{\sigma}\right)^{\alpha+1}dy_1 \nonumber\\
	&=& \boldsymbol{x}_t \boldsymbol{x}_t'   \sigma^{-\alpha-2}\int {u(s)}^2f(s)^{\alpha+1}ds
	~~~~[\mbox{using } s={(y_{1t}-\boldsymbol{\beta'x_t})}/{\sigma}, ~~~dy_1=\sigma ds] 	\nonumber\\
	&=&\sigma^{-\alpha-2} M_{f,0,2}^{(\alpha)} \boldsymbol{x}_t \boldsymbol{x}_t'  = \zeta_{11,\alpha}\sigma^{-\alpha-2}\boldsymbol{x}_t \boldsymbol{x}_t', \nonumber\\
%where $M_{f,i,j}^{(\alpha)}=\int s^i (u(s))^jf^{1+\alpha}ds$ and $\zeta_\alpha=\sigma^{-\alpha-2} M_{f,0,2}^{(\alpha)}$
\nonumber\\
B	&=&\frac{\boldsymbol{x}_t}{2\sigma^{\alpha+4}}\bigg[ \int u\left(\frac{y_{1t}-\boldsymbol{\beta'x_t}}{\sigma}\right)
		f\left(\frac{y_{1t}-\boldsymbol{\beta'x_t}}{\sigma}\right)^{\alpha+1}dy_1\nonumber\\
	&& ~~~~~~~~~~ +\int\left(\frac{ y_{1t}-\boldsymbol{\beta'x_t}}{\sigma}\right) \bigg\{u\left(\frac{y_{1t}-\boldsymbol{\beta'x_t}}{\sigma}\right)\bigg\}^2 
		f\left(\frac{y_{1t}-\boldsymbol{\beta'x_t}}{\sigma}\right)^{\alpha+1}dy_1\bigg]\nonumber\\
	&=&\frac{\boldsymbol{x}_t}{2\sigma^{\alpha+3}}\bigg[\int u(s)f(s)^{\alpha+1} ds+\int s u(s)^2f(s)^{\alpha+1} ds\bigg]\nonumber\\
	&=&\frac{\sigma^{-\alpha-3}}{2}\bigg[M_{f,0,1}^{(\alpha)}+M_{f,1,2}^{(\alpha)}\bigg]\boldsymbol{x}_t\nonumber\\
	&=&\zeta_{12,\alpha} \sigma^{-\alpha-3} \boldsymbol{x}_t, \nonumber
%where $\delta_\alpha=\frac{\sigma^{-\alpha-3}}{2}\bigg[M_{f,0,1}^{(\alpha)}+M_{f,1,2}^{(\alpha)}\bigg]$
%\nonumber\\
\end{eqnarray}
\begin{eqnarray}	
\text{and }C &=&\frac{1}{\sigma^{\alpha+5}}\bigg[\int\frac{1}{4}\left(\frac{ y_{1t}-\boldsymbol{\beta'x_t}}{\sigma}\right)^2 	
				u\left(\frac{y_{1t}-\boldsymbol{\beta'x_t}}{\sigma}\right)^2f\left(\frac{y_{1t}-\boldsymbol{\beta}'\boldsymbol{x}_t}{\sigma}\right)^{\alpha+1}dy_1
				\nonumber\\ 
 			&& ~~~~~~~~~+\int \left(\frac{y_{1t}-\boldsymbol{\beta'x_t}}{2\sigma}\right) 
 			u\left(\frac{y_{1t}-\boldsymbol{\beta'x_t}}{\sigma}\right)f\left(\frac{y_{1t}-\boldsymbol{\beta'x_t}}{\sigma}\right)^{\alpha+1}dy_1
 			+\frac{1}{4}\int f\left(\frac{y_{1t}-\boldsymbol{\beta'x_t}}{\sigma}\right)^{\alpha+1}dy_1\bigg]\nonumber\\
			&=&\frac{1}{\sigma^{\alpha+4}}\bigg[\frac{1}{4}\int s^2 u\left(s\right)^2f\left(s\right)^{\alpha+1}ds 
				+ \frac{1}{2}\int s u\left(s\right)f\left(s\right)^{\alpha+1}ds+\frac{1}{4}\int f\left(s\right)^{\alpha+1}ds\bigg]\nonumber\\
			&=&\frac{1}{\sigma^{\alpha+4}}\bigg[\frac{1}{4}M_{f,2,2}^{(\alpha)}+\frac{1}{2}M_{f,1,1}^{(\alpha)}+\frac{1}{4}M_{f,0,0}^{(\alpha)}\bigg]
			=\zeta_{22,\alpha} \sigma^{-\alpha-4}.\nonumber
\end{eqnarray}

\noindent
Thus, we get 
$\boldsymbol{J}^{(t)}=\frac{1}{\sigma^{\alpha+2}}\left(\begin{array}{cc}
	\zeta_{11,\alpha} \boldsymbol{x}_t \boldsymbol{x}_t'& \frac{\zeta_{12,\alpha}}{\sigma} \boldsymbol{x}_t\\
	\frac{\zeta_{12,\alpha}}{\sigma} \boldsymbol{x}_t'& \frac{\zeta_{22,\alpha}}{\sigma^2}
\end{array}\right)$
and, as a results,
\begin{align}
\boldsymbol{\Psi}_\alpha &=\frac{1}{T_1}\sum_{i=1}^{T_1}\boldsymbol{J}_\alpha^{(t)}
=\frac{1}{\sigma^{\alpha+2}} \left(\begin{array}{cc}
		\frac{\zeta_{11,\alpha}}{T_1} \sum_{i=1}^{T_1}\boldsymbol{x}_t \boldsymbol{x}_t' &\frac{\zeta_{11,\alpha}}{T_1\sigma}\sum_{i=1}^{T_1}\boldsymbol{x}_t\\
		\frac{\zeta_{12,\alpha} }{T_1\sigma}\sum_{i=1}^{T_1}\boldsymbol{x}_t' & \frac{\zeta_{22,\alpha}}{\sigma^2}
	\end{array}\right)
= \frac{1}{\sigma^{\alpha+2}} \left(\begin{array}{cc}
	\frac{\zeta_{11,\alpha}}{T_1} \boldsymbol{X}_0'\boldsymbol{X}_0 &\frac{\zeta_{12,\alpha}}{T_1\sigma}\boldsymbol{X}_0'\boldsymbol{1}\\
	\frac{\zeta_{12,\alpha} }{T_1\sigma}\boldsymbol{1}'\boldsymbol{X}_0 & \frac{\zeta_{22,\alpha}}{\sigma^2}
\end{array}\right).\label{3.2}
\end{align}
%where $\boldsymbol{{X_0}'}=[x_1,\cdots,x_{T_1}]$ and $\boldsymbol{1}=(1,\cdots,1)'$.\\

Next, note that, the first term in $\boldsymbol{\Omega}_\alpha$ can be derived in a similar manner as  $\boldsymbol{\Psi}_\alpha$ 
(by changing $\alpha$ to $2\alpha$) and so we are left with the computation of $\boldsymbol{\xi}_t$. 
The last element of $\boldsymbol{\xi}_t$  is given by 
\begin{align}
& -\frac{1}{2\sigma^{\alpha+3}}\int f\left(\frac{y_{1t}-\boldsymbol{\beta'x_t}}{\sigma}\right)^{\alpha+1}dy_1-\frac{1}{\sigma^{\alpha+1}}
	\int \left(\frac{y_{1t}-\boldsymbol{\beta'x_t}}{2\sigma^3}\right){u\left(\frac{y_{1t}-\boldsymbol{\beta' x_t}}{\sigma}\right)}
	f\left(\frac{y_{1t}-\boldsymbol{\beta' x_t}}{\sigma}\right)^{1+\alpha}dy_1\nonumber\\&
=-\frac{1}{2\sigma^{\alpha+2}}\bigg[\int f\left(s\right)^{\alpha+1}ds+\int s \{u\left(s\right)\}f\left(s\right)^{\alpha+1} ds\bigg]
=-\frac{1}{2\sigma^{\alpha+2}}\bigg[M_{f,0,0}^{(\alpha)}+M_{f,1,1}^{(\alpha)}\bigg]
=\sigma^{-\alpha-2}\phi_{2,\alpha}, \nonumber
\end{align}
whereas the first $N$ components of  $\boldsymbol{\xi}_t$ are given by the vector
\begin{align}
& -\frac{\boldsymbol{x}_t}{\sigma^{\alpha+2}}\int\{u\left(\frac{y_{1t}-\boldsymbol{\beta'x_t}}{\sigma}\right)\} 
	f\left(\frac{y_{1t}-\boldsymbol{\beta'x_t}}{\sigma}\right)^{\alpha+1}dy_1\nonumber\\&
=\boldsymbol{x}_t\sigma^{-\alpha-1}\int {u(s)}f(s)^{\alpha+1}ds
=\sigma^{-\alpha-1}M_{f,0,1}^{(\alpha)}\boldsymbol{x}_t 	
= \sigma^{-\alpha-1}\phi_{1,\alpha}\boldsymbol{x}_t.\nonumber
\end{align}
Thus, $ \boldsymbol{\xi}_t= \sigma^{-\alpha-1}\left(\begin{array}{c} 
									\phi_{1,\alpha}\boldsymbol{x}_t  \\
									\phi_{2,\alpha}	
								\end{array}\right)$, 
and hence
\begin{align}
\boldsymbol{\Omega}_\alpha = \frac{1}{\sigma^{2\alpha+2}} \left(\begin{array}{cc}
	\frac{[\zeta_{11,2\alpha}-\phi_{1,\alpha}^2]}{T_1} \boldsymbol{X}_0'\boldsymbol{X}_0 &\frac{[\zeta_{12,2\alpha}-\phi_{1,\alpha}\phi_{2,\alpha}]}{T_1\sigma}\boldsymbol{X}_0'\boldsymbol{1}\\
	\frac{[\zeta_{12,2\alpha}-\phi_{1,\alpha}\phi_{2,\alpha}] }{T_1\sigma}\boldsymbol{1}'\boldsymbol{X}_0 
	& \frac{[\zeta_{22,2\alpha}-\phi_{2,\alpha}^2]}{\sigma^2}
\end{array}\right). \nonumber
\end{align}
This completes the proof of our theorem by taking (in probability) limits of $\boldsymbol{\Psi}_\alpha$ and $\boldsymbol{\Omega}_\alpha$. 
\hfill{$\square$}

\section{Proof of Theorems from Section \ref{SEC:Tests}}
\label{APP:Proof_Tests}

\subsection{Proof of Theorem \ref{THM:Power_Approx1}}

%\textbf{Proof:}
\begin{align}
	\beta_{W_1^{(\alpha)}}\left(\Delta_1^*\right)&=P_{\Delta_1^*}\left(W_1^{(\alpha)}>z_\tau\right) \nonumber\\
	&=P_{\Delta_1^*}\left(\frac{\sqrt T_2(\widehat{\Delta}_1^{(\alpha)} -\Delta_{10})}{{\sqrt{\widehat{\Sigma}(\alpha)}}}>z_\tau\right)\nonumber\\
	&=P_{\Delta_1^*}\left(\frac{\sqrt T_2(\widehat{\Delta}_1^{(\alpha)} -\Delta_1^*+\Delta_1^* -\Delta_{10})}{{\sqrt{\widehat{\Sigma}(\alpha)}}}>z_\tau \right)\nonumber\\
	&=P_{\Delta_1^*}\left(\frac{\sqrt T_2(\widehat{\Delta}_1^{(\alpha)} -\Delta_1^*)}{{\sqrt{\widehat{\Sigma}(\alpha)}}}>z_\tau-\frac{\sqrt T_2(\Delta_1^*-\Delta_{10})}{{\sqrt{\widehat{\Sigma}(\alpha)}}}\right)\nonumber\\
	&\cong1-\Phi\left(z_{\tau}-\sqrt{ T_2}\frac{(\Delta^*_1-\Delta_{10})}{{\sqrt{\widehat{\Sigma}(\alpha)}}}\right).\nonumber
\end{align}
The last approximation follows from the asymptotic distribution of $\widehat{\Delta}_1^{(\alpha)}$ under the model distribution with $\Delta_1=\Delta_1^*$,
and the consistency of $\widehat{\Sigma}$.

\subsection{Proof of Theorem \ref{THM:Power_Contig1}}

%\textbf{Proof:} 
Note that 
\begin{align*}
	\sqrt{T_2}\left(\widehat{\Delta}_1^{(\alpha)} -\Delta_{10}\right)&=\sqrt{T_2}(\widehat{\Delta}_1^{(\alpha)} -\Delta_{1,T_2})-\sqrt{T_2}(\Delta_{1,T_2}-\Delta_{10})\\
	&=\sqrt{T_2}(\widehat{\Delta}_1^{(\alpha)} -\Delta_{1,T_2})-d.
\end{align*}
Under $H_{1,T_2}$, using the results from the theory of contiguity (Le Cam's third lemma), it follows that
\begin{equation*}
	\sqrt{T_2}(\widehat{\Delta}_1^{(\alpha)} -\Delta_{1,T_2})\xrightarrow{d}N(0,\Sigma(\alpha)),
\end{equation*}
and so
\begin{equation*}
	\sqrt{T_2}(\widehat{\Delta}_1^{(\alpha)} -\Delta_{10})\xrightarrow{d}N(d,\Sigma(\alpha)).
\end{equation*}
Therefore, using the consistency of $\widehat{\Sigma}$ for $\Sigma$, we have  
\begin{equation*}
	W_1^{(\alpha)}=\frac{\sqrt T_2(\widehat{\Delta}_1^{(\alpha)} -\Delta_{10})}{{\sqrt{\widehat{\Sigma}(\alpha)}}}\xrightarrow{d}N(d/\sqrt{\Sigma(\alpha)},1).  
\end{equation*}
Hence, the corresponding asymptotic power function is given by
\begin{align}
	\overline{\beta}_{W_1}^{(\alpha)}(d)&=\lim\limits_{T_2\to \infty} P_{\Delta_{1,T_2}}(W_1^{(\alpha)}>z_\tau) \nonumber\\
&= \lim\limits_{T_2\to \infty} P_{\Delta_{1,T_2}}\left(\frac{\sqrt T_2(\widehat{\Delta}_1^{(\alpha)} -\Delta_{10})}{{\sqrt{\widehat{\Sigma}(\alpha)}}} - \frac{d}{\sqrt{\Sigma(\alpha)}} 	> z_\tau - \frac{d}{\sqrt{\Sigma(\alpha)}} \right)\nonumber\\
%	&=P_{\Delta_{1,T_2}}\left(\frac{\sqrt T_2(\widehat{\Delta}_1^{(\alpha)} -\Delta_{1,T_2}+\Delta_{1,T_2} -\Delta_{10})}{{\sqrt{\widehat{\Sigma}(\alpha)}}}>z_\tau\right) \nonumber\\
%	&=P_{\Delta_{1,T_2}}\left(\frac{\sqrt T_2(\widehat{\Delta}_1^{(\alpha)} -\Delta_{1,T_2})}{{\sqrt{\widehat{\Sigma}(\alpha)}}}>z_\tau-\frac{\sqrt T_2(\Delta_{1,T_2}-\Delta_{10})}{{\sqrt{\widehat{\Sigma}(\alpha)}}}\right )\nonumber\\
	&= 1-\Phi\left(z_{\tau}-\frac{d}{{\sqrt{\Sigma(\alpha)}}}\right).\nonumber
\end{align}

\subsection{Proof of Theorem \ref{THM:Power_Approx2}}

%\textbf{Proof}
\begin{align}
	\beta_{W_2}^{(\alpha)}\left(\Delta_{11}^*,\Delta_{12}^*\right)&=P(W_2^{(\alpha)}>z_\tau) \nonumber\\
	&=P\left(\frac{(\widehat{\Delta}_{11}^{(\alpha)}-\widehat{\Delta}_{12}^{(\alpha)})}{\sqrt{\frac{\widehat{\boldsymbol{\Sigma}_1}(\alpha)}{T_{21}}+\frac{\widehat{{\Sigma}_2}(\alpha)}{T_{22}}}}>z_\tau\right) \nonumber\\
	&=P\left(\frac{(\widehat{\Delta}_{11}^{(\alpha)}-\widehat{\Delta}_{12}^{(\alpha)})}{\sqrt{\frac{\widehat{\boldsymbol{\Sigma}_1}(\alpha)}{T_{21}}+\frac{\widehat{{\Sigma}_2}(\alpha)}{T_{22}}}}-\frac{(\widehat{\Delta}^*_{11}-\widehat{\Delta}^*_{12})}{\sqrt{\frac{\widehat{\boldsymbol{\Sigma}_1}(\alpha)}{T_{21}}+\frac{\widehat{{\Sigma}_2}(\alpha)}{T_{22}}}}>z_\tau-\frac{(\widehat{\Delta}^{*}_{11}-\widehat{\Delta}^{*}_{12})}{\sqrt{\frac{\widehat{\boldsymbol{\Sigma}_1}(\alpha)}{T_{21}}+\frac{\widehat{{\Sigma}_2}(\alpha)}{T_{22}}}}\right) \nonumber\\
	&\cong 1-\Phi\left(z_{\tau}-\frac{(\widehat{\Delta}^{*}_{11}-\widehat{\Delta}^{*}_{12})}{\sqrt{\frac{\widehat{\boldsymbol{\Sigma}_1}(\alpha)}{T_{21}}+\frac{\widehat{{\Sigma}_2}(\alpha)}{T_{22}}}}\right),
\end{align}
because, under the alternative parameter values $(\Delta^*_{11},\Delta^*_{12})$, we have
$$
\left(\frac{(\widehat{\Delta}_{11}^{(\alpha)}-\widehat{\Delta}_{12}^{(\alpha)})}{\sqrt{\frac{\widehat{\boldsymbol{\Sigma}_1}(\alpha)}{T_{21}}+\frac{\widehat{{\Sigma}_2}(\alpha)}{T_{22}}}}-\frac{(\widehat{\Delta}^*_{11}-\widehat{\Delta}^*_{12})}{\sqrt{\frac{\widehat{\boldsymbol{\Sigma}_1}(\alpha)}{T_{21}}+\frac{\widehat{{\Sigma}_2}(\alpha)}{T_{22}}}}\right) \xrightarrow{d} N(0,1) \hspace{2mm} as\hspace{2mm} T_{12},T_{22} \xrightarrow{}\infty.
$$

\subsection{Proof of Theorem \ref{THM:Power_Contig2}}

%\textbf{Proof:} 
Let us first note that, at any fixed alternative value $(\Delta_{11},\Delta_{12})$, the asymptotic distributions of the Mean-MDPDE of the two ATEs yield
(along with the definition of $\eta$)
\begin{equation*}
	\sqrt{\frac{T_{21}T_{22}}{T_{21}+T_{22}}}\left\{(\widehat{\Delta}_{11}^{(\alpha)}-\widehat{\Delta}_{12}^{(\alpha)})-(\Delta_{11}-\Delta_{12})\right\}
	\xrightarrow{d}N\left(0,\eta\Sigma_1(\alpha)+(1-\eta)\Sigma_2(\alpha)\right),~~~~~\mbox{as } ~T_{21},T_{22}\to +\infty,
\end{equation*}
Then, under $H_{1,d}$, invoking Le Cam's third lemma, we get 
\begin{equation*}
	\sqrt{\frac{T_{21}T_{22}}{T_{21}+T_{22}}}\left\{(\widehat{\Delta}_{11}^{(\alpha)}-\widehat{\Delta}_{12}^{(\alpha)})-(\Delta^d_{11}-\Delta^d_{12})\right\}
	\xrightarrow{d}N\left(0,\eta\Sigma_1(\alpha)+(1-\eta)\Sigma_2(\alpha)\right),~~~~~\mbox{as } ~T_{21},T_{22}\to +\infty,
\end{equation*}
and hence
\begin{equation}
\sqrt{\frac{T_{21}T_{22}}{T_{21}+T_{22}}}\frac{(\widehat{\Delta}_{11}^{(\alpha)}-\widehat{\Delta}_{12}^{(\alpha)})}{\sqrt{\eta\Sigma_1(\alpha)+(1-\eta)\Sigma_2(\alpha)}}
\xrightarrow{d}N\left(\frac{d}{\sqrt{\eta\Sigma_1(\alpha)+(1-\eta)\Sigma_2(\alpha)}}, 1\right)~~~~~\mbox{as } ~T_{21},T_{22}\to +\infty.
\label{EQ:A1}
\end{equation}

\noindent
Further, by the consistency of $\widehat{\Sigma}_j(\alpha)$ for $\Sigma_j(\alpha)$, $j=1,2$, and the definition of $\eta$,  we have
\begin{equation}
	\frac{T_{22}\widehat{\boldsymbol{\Sigma}_1}(\alpha)+ T_{21}\widehat{{\Sigma}_2}(\alpha)}{T_{21}+T_{22}}
	\xrightarrow{\mathcal{P}} \eta\Sigma_1(\alpha)+(1-\eta)\Sigma_2(\alpha),~~~~~\mbox{as } ~T_{21},T_{22}\to +\infty. 
	\label{EQ:A2}
\end{equation} 

\noindent
Therefore, combining \eqref{EQ:A1} and \eqref{EQ:A2} via Slutsky's theorem, under $H_{1,d}$, we get 
\begin{eqnarray}
W_2^{(\alpha)}&=&\frac{(\widehat{\Delta}_{11}^{(\alpha)}-\widehat{\Delta}_{12}^{(\alpha)})}{
	\sqrt{\frac{\widehat{\boldsymbol{\Sigma}_1}(\alpha)}{T_{21}}+\frac{\widehat{{\Sigma}_2}(\alpha)}{T_{22}}}}
\nonumber\\
&=& \sqrt{\frac{\eta\Sigma_1(\alpha)+(1-\eta)\Sigma_2(\alpha)}{
	\frac{T_{22}\widehat{\boldsymbol{\Sigma}_1}(\alpha)+ T_{21}\widehat{{\Sigma}_2}(\alpha)}{T_{21}+T_{22}}}} \times 
 \sqrt{\frac{T_{21}T_{22}}{T_{21}+T_{22}}}\frac{(\widehat{\Delta}_{11}^{(\alpha)}-\widehat{\Delta}_{12}^{(\alpha)})}{\sqrt{\eta\Sigma_1(\alpha)+(1-\eta)\Sigma_2(\alpha)}}
\nonumber\\
&& \xrightarrow{d} N\left(\frac{d}{\sqrt{\eta\Sigma_1(\alpha)+(1-\eta)\Sigma_2(\alpha)}}, 1\right),
~~~~~\mbox{as } ~T_{21},T_{22}\to +\infty. 
\end{eqnarray}
Hence, the corresponding asymptotic power function is  given by
\begin{align}
\overline{\beta}_{W_2}^{(\alpha)}\left(d\right) &= \lim_{T_{21},T_{22}\to +\infty} P\left(W_2^{(\alpha)}>z_\tau\right) \nonumber\\
%	&=P\left(\frac{(\widehat{\Delta}_{11}^{(\alpha)}-\widehat{\Delta}_{12}^{(\alpha)})}{\sqrt{\frac{\widehat{\boldsymbol{\Sigma}_1}(\alpha)}{T_{21}}+\frac{\widehat{{\Sigma}_2}(\alpha)}{T_{22}}}}>z_\tau\right)\nonumber\\
%	&=P\left(\frac{(\widehat{\Delta}_{11}^{(\alpha)}-\widehat{\Delta}_{12}^{(\alpha)})-(\Delta_{11}-\Delta_{12}-\sqrt{\frac{T_{21}+T_{22}}{T_{21}T_{22}}}d)+(\Delta_{11}-\Delta_{12}+\sqrt{\frac{T_{21}+T_{22}}{T_{21}T_{22}}}d)-(\Delta_{11}-\Delta_{12})}{\sqrt{\frac{\widehat{\boldsymbol{\Sigma}_1}(\alpha)}{T_{21}}+\frac{\widehat{{\Sigma}_2}(\alpha)}{T_{22}}}}>z_{\tau}\right)\nonumber\\
%	&=P\left(\frac{(\widehat{\Delta}_{11}^{(\alpha)}-\widehat{\Delta}_{12}^{(\alpha)})-(\Delta_{11}-\Delta_{12}-\sqrt{\frac{T_{21}+T_{22}}{T_{21}T_{22}}}d)}{\sqrt{\frac{\widehat{\boldsymbol{\Sigma}_1}(\alpha)}{T_{21}}+\frac{\widehat{{\Sigma}_2}(\alpha)}{T_{22}}}}>z_\tau-\frac{(\Delta_{11}-\Delta_{12}-\sqrt{\frac{T_{21}+T_{22}}{T_{21}T_{22}}}d)-(\Delta_{11}-\Delta_{12})}{\sqrt{\frac{\widehat{\boldsymbol{\Sigma}_1}(\alpha)}{T_{21}}+\frac{\widehat{{\Sigma}_2}(\alpha)}{T_{22}}}}\right) \nonumber\\
 &= \lim_{T_{21},T_{22}\to +\infty} P\left(\frac{(\widehat{\Delta}_{11}^{(\alpha)}-\widehat{\Delta}_{12}^{(\alpha)})}{
 	\sqrt{\frac{\widehat{\boldsymbol{\Sigma}_1}(\alpha)}{T_{21}}+\frac{\widehat{{\Sigma}_2}(\alpha)}{T_{22}}}} 
 -\frac{d}{\sqrt{\eta\Sigma_1(\alpha)+(1-\eta)\Sigma_2(\alpha)}} > z_\tau -\frac{d}{\sqrt{\eta\Sigma_1(\alpha)+(1-\eta)\Sigma_2(\alpha)}}\right) \nonumber\\
&= 1-\Phi\left(z_{\tau}-\frac{d}{\sqrt{\eta\Sigma_1(\alpha)+(1-\eta)\Sigma_2(\alpha)}}\right).\nonumber
\end{align}

%\newpage

\end{document}